\documentclass[11pt,letter]{article}
\usepackage{graphicx}
\usepackage{amsmath}
\usepackage{amssymb}
\usepackage{cancel}
\usepackage{url}
\usepackage[left=0.8in,right=0.8in,top=0.75in,bottom=0.75in]{geometry}
\usepackage{setspace}
\usepackage{float}
\usepackage[super,sort&compress,comma]{natbib}
\usepackage[font={small}]{caption} %scriptsize = 8pt; footnotesize = 9pt; small = 10pt; normalsize = 11pt (set on line 4 above)
\usepackage[labelfont=bf]{caption}
\usepackage{bm}
\usepackage[T1]{fontenc}
\usepackage{microtype}
\usepackage{subfigure}
\usepackage{chemformula}
\usepackage[version=4]{mhchem}
\usepackage{textcomp}
\usepackage{color,soul}
\usepackage{mathtools}
\usepackage{bm}
\usepackage{gensymb}
\usepackage{esvect}
\usepackage{booktabs}
\usepackage{multirow}
\usepackage{physics}
\usepackage{mathrsfs}
\usepackage{booktabs}
\usepackage{longtable}

\usepackage{times}
\usepackage{url}
\usepackage{algorithm}
\usepackage{algorithmic}
\usepackage{latexsym}
\usepackage{multirow}
\usepackage{array}
\usepackage{graphicx}
\usepackage{geometry}
\usepackage{float}
\usepackage{subfigure}
\usepackage{enumerate}
\usepackage{xcolor}
\usepackage{amsmath}
\usepackage{caption}
\usepackage{diagbox}
\usepackage{chemfig}
\usepackage{xurl}

\usepackage{listings}
\begin{document}

\title{\textbf{SmileyLlama: Modifying Large Language Models\\for Directed Chemical Space Exploration}}
\author{Joseph M. Cavanagh$^{1}$, Kunyang Sun$^{1}$, Andrew Gritsevskiy$^{2}$,\\ Dorian Bagni$^{1}$, Yingze Wang$^{1}$, Thomas D. Bannister$^{3}$,\\ 
Teresa Head-Gordon$^{1,4,5\dagger}$}
\date{}
\maketitle

\begin{center}
$^1$Kenneth S. Pitzer Theory Center and Department of Chemistry, \\
University of California, Berkeley, CA, 94720 USA \\
$^2$RunRL, 414 Gough St, Suite 2, San Francisco CA 94102 \\
%$^2$Contramont Research, San Francisco, CA, 94158 USA \\
$^3$Department of Molecular Medicine, The Herbert Wertheim UF Scripps Institute for Biomedical Innovation and Technology, 130 Scripps Way, Jupiter, FL, 33458 USA \\
$^4$Departments of Bioengineering and Chemical and Biomolecular Engineering, \\
University of California, Berkeley, CA, 94720 USA \\
$^5$Chemical Sciences Division, Lawrence Berkeley National Laboratory, Berkeley, CA, 94720 USA \\
$\dagger$ thg@berkeley.edu
\end{center}

\begin{abstract}
\noindent
We show that large language model (LLMs) can be transformed via supervised fine-tuning (SFT) of engineered prompts into SmileyLlama for exploring the chemical space of drug molecules. We benchmark SmileyLlama against pre-trained LLMs and chemical language models (CLM) trained from scratch for generating valid and novel drug-like molecules, and use direct preference optimization (DPO) to both improve SmileyLlama's adherence to a prompt and as part of the iMiner reinforcement learning framework to predict molecules with optimized 3D conformations and high binding affinity to drug targets. By training an LLM to speak directly as a CLM, while retaining most of its natural language capabilities, we show that we can reliably generate molecules with user-specified properties rather than acting only as a chatbot with knowledge of chemistry or as a virtual assistant. While SmileyLlama is geared toward drug discovery, the SFT/DPO/LLM framework can be extended to other chemical, biological, and materials applications. 
\end{abstract}

% the * after section prevents numbering
\section{Introduction}
Chemical Language Models (CLMs)\cite{grisoni} trained on string representations such as Simplified Molecular-Input Line-Entry System (SMILES)\cite{smiles} and SELF-referencIng Embedded Strings (SELFIES)\cite{selfies} have emerged as a useful tool for \textit{de novo} generation of molecules, best exemplified by molecules relevant to pharmaceutical applications and drug discovery\cite{mol-gen-bg2, mol-gen-bg3, mol-gen-bg4}. Nearly all CLMs for molecular generation have been trained from scratch on large quantities of data such as ChEMBL\cite{gaulton_ChEMBL_2012} and ZINC\cite{Tingle2023} and using different model architectures, including variational autoencoders\cite{VAE2018}, recurrent neural networks (RNNs)\cite{mol-gen-rnn}, generative pre-trained transformers (GPTs)\cite{gpt} and structured state space sequence (S4) models\cite{Ozcelik2024s4}. In addition, CLMs have achieved advances in chemical generation tasks through further downstream optimization of molecules with additional training or different model frameworks.\cite{mol-gen-gpt1,iminer} 

Language models are statistical models of probability distributions of units of language, and can be adapted to generate meaningful text by sampling from these distributions. The most recent advancement of language models have resulted from the training of scaled-up transformers \cite{vaswani2023attentionneed} on massive amounts of data, resulting in the creation of Large Language Models (LLMs)  such as the proprietary GPT-4\cite{openai_gpt-4_2024} and the open-weight Llama\cite{dubey_llama_2024}. Recently scientific groups have accessed frontier LLMs for the purpose of assisting research in the form of virtual lab members, to translate between natural and chemical languages\cite{yu2024llasmol},  or even performing research autonomously.\cite{Zheng2023,boiko_autonomous_2023, m_bran_augmenting_2024} Beyond chemical
dictionary lookups or lab aides, LLMs have been used to perform mutation and crossover for an evolutionary algorithm to explore chemical space\cite{molleo} or to modify SMILES strings to change the properties of the molecules that they represent.\cite{guevorguian_small_2024,drugllm} Others have taken inspiration from LLMs to design CLMs, such as the ability to respond to prompts through a transformer-based architecture\cite{wu_leveraging_2024}. However, to our knowledge, no CLM derived from a pre-trained general purpose LLM has reached the performance exhibited by modern CLMs that are trained from scratch with chemical data. 

Here, we demonstrate that an open-weight LLM, Meta-Llama-3.1-8B-Instruct (``Llama'' here on out)~\citep{dubey_llama_2024}, can be converted into a model for generative tasks in drug discovery. The fact that Llama is open-weight offers several benefits including allowing training and sharing adapters, to perform inference without needing to store potentially valuable data on a remote server, to have control over the hyperparameters and algorithms used for fine-tuning, and to perform interpretability analyses on the model weights. Using supervised fine-tuning (SFT) and Direct Preference Optimization (DPO)\cite{rafailov_direct_2024} of the pre-trained Llama with SMILES strings derived from ChEMBL, SmileyLlama generates drug-like molecules with desirable properties specified in a user-defined prompt with relevance to medicinal chemistry, which we show can match or exceed the performance of modern CLMs. We further demonstrate that SmileyLlama greatly improves the reinforcement learning component of iMiner algorithm\cite{iminer} to more efficiently explore chemical space to create molecules optimized for 3D binding to target proteins, illustrated with the SARS-Cov-2 (SARS2) main protease (MPro).\cite{mpro_review} While our dataset and subsequent analyses are created with drug discovery as a downstream application, this general procedure can be extended to other chemical applications such as chemical synthesis planning\cite{Sun2025} or transition metal complex discovery\cite{tmc2026}.

\section{Results}

\subsection{Supervised Fine-Tuning and Direct Preference Optimization of Llama}
In order to steer the outputs of the pre-trained Llama model~\citep{dubey_llama_2024} for drug molecule generation we first use supervised fine tuning (SFT), in which the weights of Llama are further optimized on SMILES strings of approximately 2 million molecules from the ChEMBL Datasetv33 \cite{gaulton_ChEMBL_2012} to create SmileyLlama. For each molecule in our dataset, we picked a number of molecular properties of pharmaceutical interest to calculate using RDKit\cite{Landrum2016RDKit2016_09_4} and that are relevant for medicinal chemistry. In addition, drug molecules must also have suitable characteristics related to relevant biological phenomena such as obeying the Rule of Five\cite{jhoti_rule_2013}, or TPSA ranges which tend toward being orally bioavailable or able to pass through the placenta or blood-brain-barrier\cite{veber_molecular_2002}. If a drug need not meet these criteria then a user interfacing with SmileyLlama should also be able to adjust the range criterion or eliminate it. Further specifics of these properties and the ranges we choose to specify during training of SmileyLlama can be found in the Methods Section 4.1. 

After calculating and picking these properties for each SMILES string, we construct a prompt containing values of these properties, with the ``correct'' completion being the SMILES string that these properties were calculated from. To illustrate, we used a prompt with a system instruction of \textbf{You love and excel at generating SMILES strings of drug-like molecules} and a user instruction of the form \textbf{Output a SMILES string for a drug like molecule with the following properties:} if properties are specified, or \textbf{Output a SMILES string for a drug like molecule:} if no properties are specified. We choose to create prompts that assign SmileyLlama the role of an AI which excels at producing SMILES strings due to the track record of role-prompting\cite{chenUnleashingPotentialPrompt2025}; we also chose this prompt format due to its balance between motivation and brevity. Each property has a 50\% chance of being calculated and specified in the prompt so that the trained model learns to operate equally well during inference, whether or not any properties are specified. We structure the prompts used for SFT so that during inference users avoid having to downselect the vast majority of generated molecules for having the correct characteristics—instead, users can simply prompt SmileyLlama to provide molecules with the characteristics they desire. See Methods Section 4.2 and 4.3, as well as Supplementary Algorithm 1 and Supplementary Figure 1 for further elaboration of SFT training. 

We also use Direct Preference Optimization (DPO)\cite{rafailov_direct_2024} which also updates the weights of Llama to  reinforce our model's ability to robustly generate molecules for more specific task-oriented goals such as property specification. Algorithmically we prompt our SFT model to generate molecules with a given property, and sample several SMILES strings and use RDKit\cite{Landrum2016RDKit2016_09_4} to assess whether they have properties in line with what the prompt requested. We then pair up molecules which correctly followed the prompt with those that don't as winners and losers, respectively, then use a single epoch of DPO to improve the model's results. See Supplementary Algorithm 2 for pseudocode of this scoring and pairing procedure. 

\subsection{Benchmarking SmileyLlama against other LLMs and CLMs}
To test the generative ability of SmileyLlama compared with other existing CLMs, we used the GuacaMol suite~\cite{brown2019guacamol} to benchmark the validity, uniqueness, and novelty of the molecules as shown in Table \ref{tab:guacamol}. Additionally, KL divergence and Frechet Chemnet Distance (FCD)~\cite{preuer2018frechet} based on the GuacaMol definition (FCD$_{Guac}$) are used to analyze the distributional shifts from the ChEMBL training data for drug-like molecules.\cite{brown2019guacamol}.  More detail is found in Methods Section 4.4.

We first analyze the ability of Llama to produce molecules, relying only on its pre-trained knowledge (zero-shot), or by providing it with one or more examples from the ChEMBL database in the formulated prompt (Table \ref{tab:guacamol} and Supplementary Table 1). We find that without SFT or examples provided in a prompt the LLM is unable to produce high percentages of valid SMILES strings compared to other state-of-the-art CLMs, and is generally poor even with variations in hyperparameters such as temperature ($T$). Interestingly, validity is lower when 20 examples are provided in the prompt (twenty-shot) than it is when no examples are in the prompt (zero-shot). We speculate that Llama zero-shot  has had some exposure to the SMILES syntax to be able to generate valid strings, but it has no intrinsic ability to generalize, repeating the memorized SMILES and resulting in low uniqueness. When several examples are given in the prompt, this biases Llama away from the known SMILES strings it can produce,  but the few examples means that its grasp on the allowed mutable structure of SMILES strings is poor and thus less valid. However, because these prompts are so diverse, Llama's 20-shot uniqueness is very high. 

\begin{table}[h]
    \centering
    \caption{\textit{GuacaMol benchmarks comparing SmileyLlama to LLMs and to common CLM architectures trained on ChEMBL.} The model benchmarks include valid chemical molecules, uniqueness and novelty with respect to the training set, and distribution similarity evaluated using KL divergence and Frechet ChemNet distance based on the GaucaMol definition FCD$_{Guac}=exp(-0.2*FCD)$. Additional information is available in Methods Section 4.4.}
    \small
    \vspace{-2mm}
    \begin{tabular}{lcccccc}
        \hline\hline
        \textbf{Benchmark} & \textbf{Validity} & \textbf{Uniqueness} & \textbf{Novelty} & \textbf{KL div} & \textbf{FCD$_{Guac}$} \\ \hline
        \textbf{GraphMCTS\cite{Iwata2023}} & 1.000 & 1.000 & 0.994 &  0.522 & 0.015  \\
        \textbf{VGAE-MCTS\cite{Iwata2023}} & 1.000 & 1.000 & 1.000 &  0.659 & 0.009  \\
        \textbf{AAE\cite{Iwata2023}} & 0.822 & 1.000 & 0.998 &  0.886 & 0.529  \\
        \textbf{LSTM\cite{Ozcelik2024s4}} & 0.983 & 0.999 & 0.848 & 0.993 & 0.901 \\
        \textbf{GPT\cite{Ozcelik2024s4}} & 0.915 & 1.000 & 0.978 & 0.977 & 0.826 \\
        \textbf{S4 \cite{Ozcelik2024s4}} & 0.971 & 0.997 & 0.961 & 0.994 & 0.853 \\        
        \textbf{Llama zero-shot} & 0.688 & 0.457 & 0.635 & 0.736 & 0.002 \\
        \textbf{Llama twenty-shot} & 0.465 & 0.999 & 0.949 & 0.913 & 0.079 \\
        \textbf{SmileyLlama} & 0.958 & 1.000 & 0.987 & 0.967 & 0.686 \\  %\textbf{SmileyLlama} & 0.960 & 1.000 & 0.988 & 0.970 & 0.553 \\
        \hline
    \end{tabular}
\label{tab:guacamol}
\end{table}

In Table \ref{tab:guacamol} it is seen that SFT substantialy improves SmileyLlama’s ability to generate drug-like molecules. In addition, we experiment with the format of the SmileyLlama prompt, performing SFT on Llama with a less anthropomorphic user prompt and a blank template as an ablation study, showing that changing this prompt format does not substantially affect the Guacamol benchmarks (Supplementary Table 1). To show the generality of the LLM-SFT approach, we also fine-tune Llama-3.2-3B, Llama-3.2-1B, and Qwen-2.5-7B\cite{qwen25TechnicalReport2025} using the same SFT workflow (including identical hyperparameters) that we developed for SmileyLlama. Supplementary Table 1 finds that the Guacamol benchmark results did not change substantially between SmileyLlama and SmileyQwen2.5-7B. We also find through inspecting SmileyLlama-1B and SmileyLlama-3B, that the validity increases with parameter count, though novelty, uniqueness and the match between the training distribution and the distribution of generated molecules stays largely unchanged. 

\begin{figure}[H]
\centering
\includegraphics[width=0.9\textwidth]{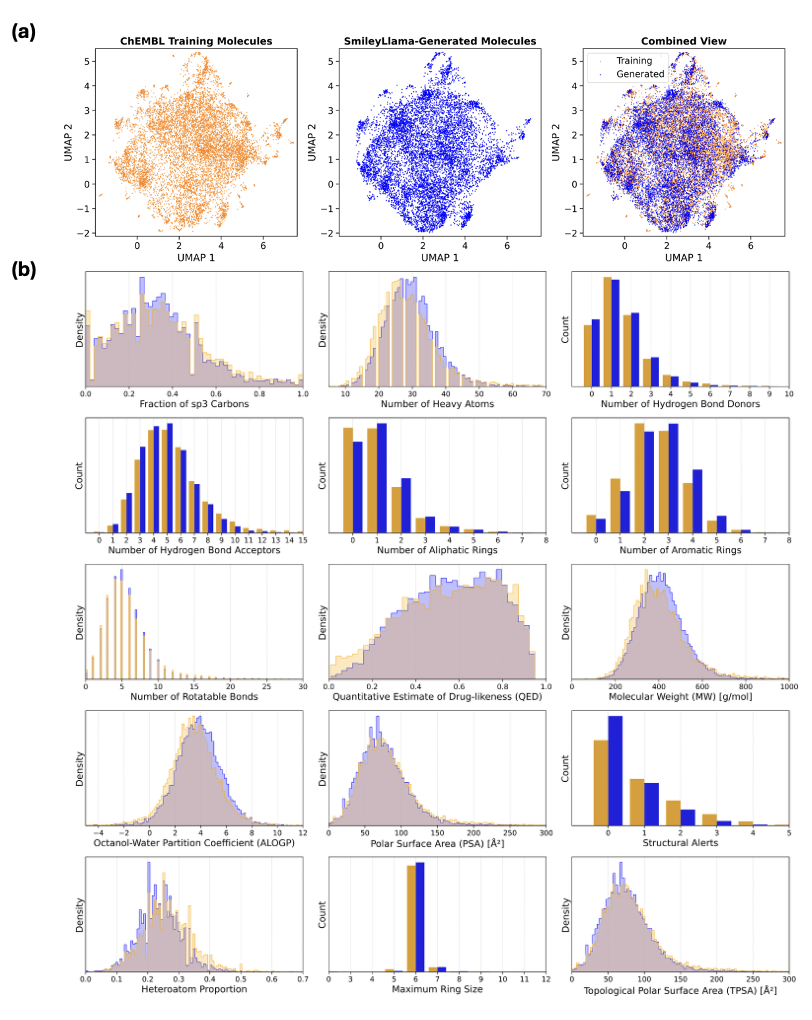}
\vspace{-3mm}
\caption{\textit{Distribution comparisons for different properties of the generated molecules from SmileyLlama (blue) with molecules from the training dataset from ChEMBL (gold).} (a) UMAP visualization of a random selection of 10,000 ChEMBL molecules and 10,000 SmileyLlama-generated molecules, using 15 neighbors and a minimum distance of 0.1; these are normal values in chemical space visualization\cite{orlovHighDimensionsHuman2025}. (b) The molecular properties considered are fraction of $sp^3$ hybridized carbons and heteroatoms, number of heavy atoms, number of H-bond donors and acceptors, number of aliphatic and aromatic rings and the maximum ring size, number of rotatable bonds, quantitative estimate of drug-likelihood (QED) value, MW, approximate log partition coefficient between octanol and water (ALOGP), polarizable surface area (PSA) and topological PSA, and the number of structural alerts. All benchmarks were at a temperature $T=$1.0 and a maximum of 256 new tokens.}
\label{fig:dist}
\end{figure}

Figure \ref{fig:dist} shows that SmileyLlama generates very good agreement with ChEMBL quantities across a diverse property set. The UMAP visualization in Figure \ref{fig:dist}a, a popular visualization tool used in drug discovery, finds that

SmileyLlama generates molecules in every well-represented region of the chemical space of ChEMBL. We also consider the distribution of molecular properties of interest to medicinal chemistry in Figure \ref{fig:dist}b, in which the KL-divergence values indicate all properties are in very good agreement between the SmileyLlama-generated molecules and ChEMBL molecules, and comparable to that from other models with high KL-divergence from Guacamole (Table \ref{tab:guacamol} and Supplementary Figures 3 and 4). Furthermore, small percentages of undesirable molecular scaffolds are present in the ChEMBL training data itself\cite{iminer}, but Supplementary Table 2 shows that SmileyLlama and most robust CLMs do not oversample these unviable chemical structures. Finally, while training was conducted at $T=1.0$, exploration of temperature used at inference on the Guacamol benchmark (Supplementary Figure 2) suggests that this temperature is adequate for all tests described in the Results.

\subsection{Property Specification using SmileyLlama under SFT}
In Table \ref{tab:properties_specification} we show the average percentage of valid, distinct SMILES strings generated for a complete panel of of molecular property tasks with SFT. This benchmark is distinct from other conditional molecule generation benchmarks\cite{yu2024llasmol} in that we are testing SmileyLlama's ability to robustly generate molecules with properties in value ranges rather than a specific value. This is of interest to medicinal chemists where numerical ranges of Lipinski violations or hydrogen bond donors and acceptors (and others) are used during chemical exploration. Additionally, LLMs tend to struggle with numbers which have many degrees of precision and must be split into several tokens\cite{schwartz-etal-2024-numerologic}. Hence we did not represent this category in the prompt during training. 

\begin{table}[H]
    \centering
    \caption{\textit{Percentage of valid, distinct generated molecules over a panel of tasks using SmileyLlama}. For each task, we generate 1000 molecules from a prompt requesting some property, and score the result based on the proportion of molecules which are valid, distinct, and satisfy the properties requested in the prompt. Finally, we collect some tasks into families (such as those with different range values) and average their scores to produce the results shown in the table below. All benchmarks were at a temperature $T=$1.0 and a maximum of 128 new tokens to increase computational efficiency. We also note that running SFT for two epochs rather than one did not seem to greatly affect the performance on these benchmarks.}
    \vspace{-2mm}
    \small
    \begin{tabular}{lcccccc}
        \hline\hline
        & \multicolumn{2}{c}{SFT} & \multicolumn{2}{c}{DPO}  & \multicolumn{2}{c}{Prompt Ablation}\\
        \textbf{Present in Trained Prompt} & T=0.7 & T=1.0 & T=0.7 & T=1.0 & T=0.7 & T=1.0 \\
        \hline
$\leq$ k H-bond donors              & 96.6\% & 94.4\% & 99.2\% & 98.1 \% & 93.8\% & 91.8\%\\
$\leq$ k H-bond acceptors           & 96.3\% & 90.7\% & 98.3\% & 98.4 \% & 76.6\% & 65.4\%\\
$\leq$ k Molecular weight           & 90.1\% & 84.3\% & 97.8\% & 98.0 \% & 62.4\% & 58.8\%\\
$\leq$ k ClogP                      & 89.2\% & 85.4\% & 96.8\% & 97.7 \% & 67.9\% & 64.9\%\\
Rotatable bonds in range            & 90.0\% & 86.7\% & 94.5\% & 94.0 \% & 62.2\% & 58.5\%\\
Fraction $sp^3$ in range            & 87.6\% & 85.0\% & 96.3\% & 95.9 \% & 30.7\% & 31.8\%\\
TPSA in range                       & 95.9\% & 91.1\% & 98.8\% & 98.4 \% & 89.2\% & 83.1\%\\
No bad SMARTS                       & 92.5\% & 89.6\% & 94.9\% & 94.3 \% & 87.3\% & 85.6\%\\
Absence of macrocycle               & 98.4\% & 95.3\% & 98.4\% & 97.3 \% & 97.1\% & 94.8\%\\
Absence of warhead-related SMARTS   & 96.4\% & 94.2\% & 97.3\% & 95.4 \% & 94.6\% & 92.9\%\\
Lipinski rule-of-five               & 89.0\% & 81.7\% & 98.7\% & 97.7 \% & 73.6\% & 64.0\%\\
Presence of warhead-related SMARTS  & 58.3\% & 51.0\% & 74.9\% & 73.0 \% & 0.2\%  & 0.4\%\\
Presence of Enamine substructures   & 51.0\% & 51.5\% & 67.7\% & 70.1 \% & 2.7\%  & 1.9\%\\
Presence of macrocycle              & 38.8\% & 44.7\% & 58.0\% & 57.8 \% & 2.1\%  & 2.1\%\\
\hline\hline
        \textbf{Absent in Trained Prompt} & T=0.7 & T=1.0 & T=0.7 & T=1.0 & T=0.7 & T=1.0 \\
        \hline
Rule-of-three                       & 77.5\% & 62.0\% & 84.8\% & 94.5\% & 7.5\% & 5.5\%\\
Exactly k H-bond donors             & 21.4\% & 19.7\% & 30.7\% & 30.4\% & 17.4\% & 16.8\%\\
Exactly k H-bond acceptors          & 19.9\% & 14.4\% & 27.0\% & 31.9\% & 9.2\% & 8.8\%\\
    \end{tabular}
\label{tab:properties_specification}
\end{table}

Overall SmileyLlama model does very well on tasks on which it was trained through the engineered prompt, especially when contrasted with the model resulting from the "prompt ablation" experiment in Table \ref{tab:properties_specification}. We note that one has a choice to use SmileyLlama using lower temperatures at inference that can improve the SFT predictions further. Although all individual properties were present in training, some were represented less frequently, such as the Lipinski Rule of-5, the presence of a macrocycle, and some categories of warhead-related SMARTS and  
\vspace{-3mm}
\begin{figure}[H]
\centering
\includegraphics[width=0.9\textwidth]{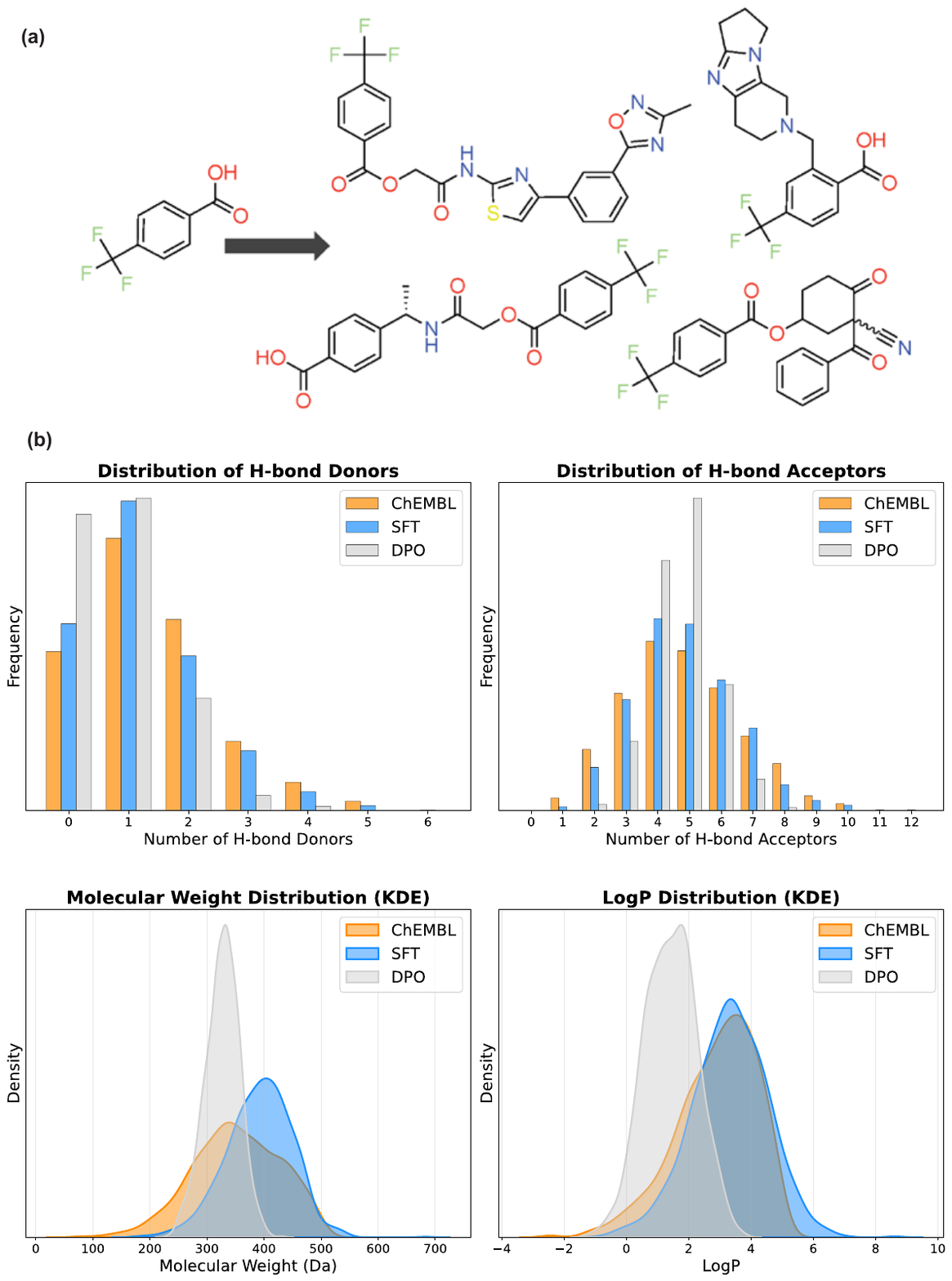}
\vspace{-3mm}
\caption{\textit{Conditional generation with SmileyLlama for fragment growth and before and after DPO compared to ChEMBL.} (a) Example molecules generated by growing from one of the Enamine substructures and to satisfy Lipinski's Rule of 5 using the prompt \textbf{Output a SMILES string for a drug like molecule with the following properties: a substructure of O=C(O)c1ccc(C(F)(F)F)cc1, <= 500 MW, <=5 logP, <= 5 H-bond donors, <= 10 H-bond acceptors}. (b) distribution of four properties satisfying Lipinski's rule of five comparing ChEMBL molecules (orange) with molecules generated by SmileyLlama (blue) with the prompt \textbf{Output a SMILES string for a drug like molecule with the following properties: <= 5 H-bond donors, <= 10 H-bond acceptors, <= 500 MW, <= 5 logP}, compared to 1000 molecules generated by SmileyLlama with the same prompt after DPO (gray). MW and LogP distributions were estimated using a gaussian kernel density estimator (KDE). All results generated 1000 molecules at a temperature $T=$1.0 and a maximum of 128 new tokens.}
\label{fig:dpo_sft}
\end{figure}
\noindent
Enamine substructures, resulting in performance that is more middling for these categories. We compare SmileyLlama with a model resulting from an ablation study on the efficacy of prompting.  As expected, SmileyLlama does poorly on tasks involving exact numerical specifications. More encouragingly, SmileyLlama performs well on compound tasks such as generating molecules similar to existing leads, i.e. “scaffold hopping” R-group modification, and/or structure-based design to grow molecules from ligand fragments. Figure \ref{fig:dpo_sft}a is an example of SmileyLlama model's ability to generate molecules from all 320 substructures in the Enamine database\cite{enamine_essential_libraries} that follow the Lipinski rule-of-five\cite{LIPINSKI20013}, which encompasses most of the molecular properties with ranges listed in Table 2. 

We study this by removing all indications of molecular properties from all of the prompts in the dataset used to train SmileyLlama; each molecule from ChEMBL is a completion to the same prompt (the prompt used for SmileyLlama when no properties of a molecule are selected). When we run SFT with exactly the same hyperparameters as SmileyLlama, we find that the ablated model performs quite poorly in comparison to SmileyLlama on this benchmark, achieving 90+\% performance on only three tasks. This becomes especially pronounced when the properties are rarely found in the data, such as the presence of a macrocycle or a warhead-related SMARTS pattern. The stark contrast in performances speaks to the necessity of our prompt engineering scheme: we cannot rely purely on the knowledge of the foundation model when fine-tuning for chemical tasks.

\subsection{Property Specification using SmileyLlama under DPO}
While SmileyLlama typically does well on tasks on which it was trained through the engineered prompt, and can still perform adequately when queried with prompts different from ones on which it was trained, we can further optimize SmileyLlama for specific tasks using DPO. DPO's most popular application has been in improving the responses of LLM-derived chatbots, but it has also found use in improving the outputs of CLMs~\cite{park_preference_2023} and avoiding the need to separately train a reward model~\cite{rafailov_direct_2024}. Here the relevance of DPO provides a way to further optimize the model by pairing desirable responses with undesirable responses. The model's weights are then updated to be more likely to produce the `winner' of the pairing and less likely to produce the `loser' of the pairing. We generated our dataset by simply pairing up unsuccessful attempts at generating structures with successful attempts randomly for each task in Table 2.

SmileyLlama optimized with DPO substantially improved adherence to the prompt across nearly all tasks as seen in Table \ref{tab:properties_specification} and Figure \ref{fig:dpo_sft}b. Note that while DPO does cause the model to more robustly obey the rules in the prompt, it also shifts and narrows the property distribution compared to the training set and appears to be largely insensitive to temperature. SmileyLlama without DPO, on the other hand, occasionally does not obey the prompt but more faithfully reproduces the distribution of properties found in a filtered ChEMBL that satisfy Lipinski's rules. In the context of drug discovery, SFT is primarily beneficial for early exploration of chemical space, whereas DPO is a type of constraint optimization that limits generated molecules to desired sub-classes specified by the user.

\subsection{Binding Affinity to Protein Active Sites with SmileyLlama/iMiner}
\noindent
The tests performed in previous Sections do not take advantage of the 3-dimensional structural information of a putative drug nor its shape and molecular compatibility with a target protein active site. Hence we use SmileyLlama augmented with DPO to generate unique and valid ligands which undergo further optimization for binding to a specific protein target when embedded in the iMiner framework\cite{iminer}. iMiner combined with SmileyLlama is designed to generate novel inhibitor molecules for target proteins by combining deep reinforcement  learning\cite{olivecrona2017molecular,popova2018deep} with real-time 3D molecular docking using AutoDock Vina\cite{Trott2010}, thereby simultaneously creating chemical novelty while constraining molecules for shape and molecular compatibility with target active sites. Further details of the iMiner reinforcement learning model have been published elsewhere \cite{iminer} and are briefly summarized in Methods Section 4.5. To validate the effectiveness of SmileyLlama in the iMiner context, we generate inhibitor molecules for MPro, an enzyme whose function is essential to the SARS-CoV-2 lifecycle\cite{jin2020structure}. MPro has readily available experimental 3D structures\cite{jin2020structure,zhang2020crystal} which provides the necessary information needed for structure-based ligand design. 

For the unconditional \textit{de novo} generation case, SmileyLlama learns the user prompt \textbf{Output a SMILES string for a drug like molecule with the following properties: High SARS2PRO} which pertains to minimizing the AutoDock Vina score while maximizing the drug likeliness score ($S_{DL}$) of the original iMiner reward function\cite{iminer}. 
\begin{figure}[H]
\centering
\includegraphics[width=0.95\textwidth]{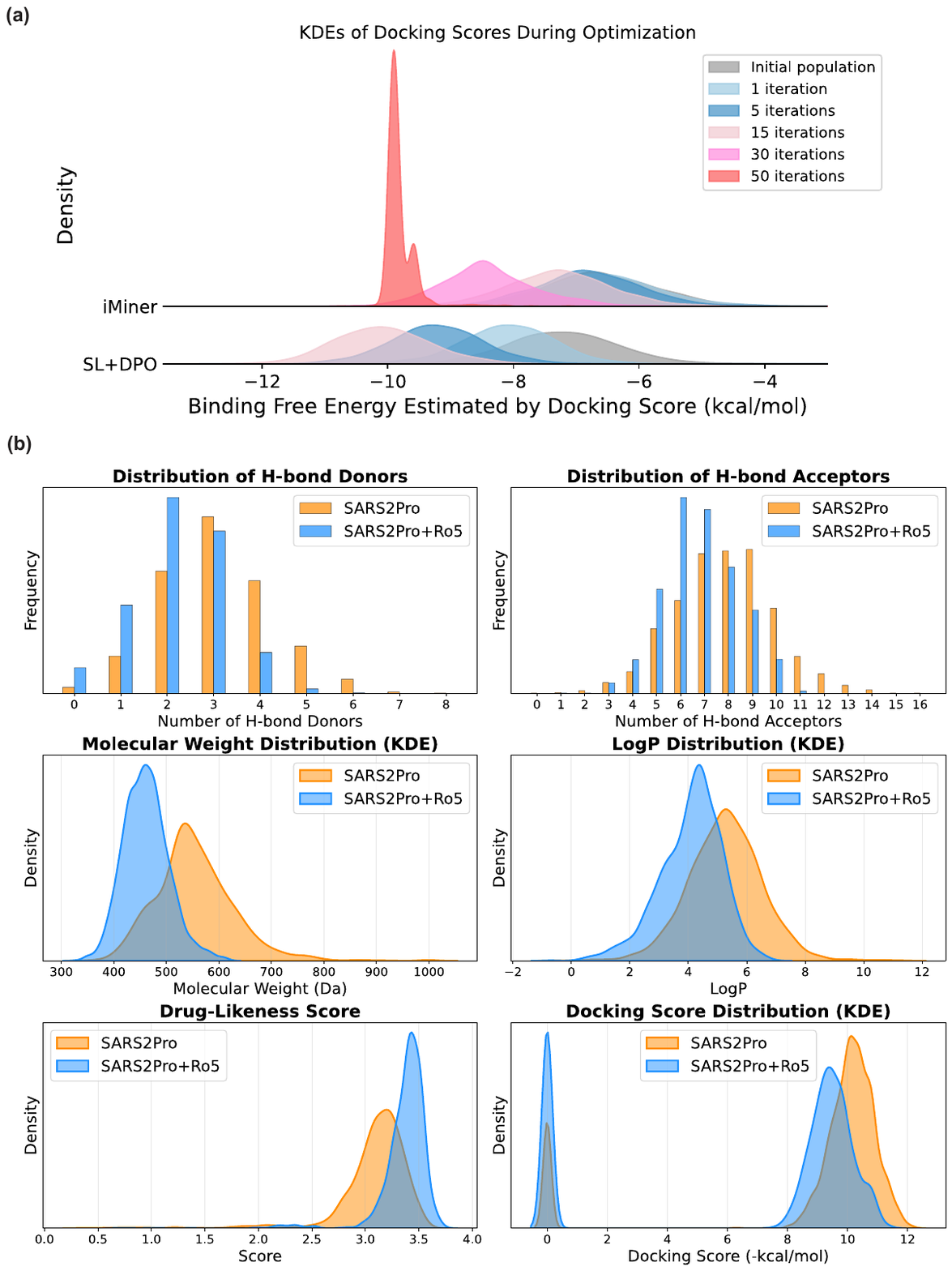}
\vspace{-3mm}
\caption{\textit{Comparison of the shift in docking score distributions for iMiner compared to SmileyLlama over optimization epochs as illustrated for SARS2-MPro.} (a) For iMiner, in later epochs diversity crashes which explains the sharpening peaks in later iterations. SmileyLlama with SL+DPO enforces diversity throughout the optimizations, which accounts for the broad peaks, and shows superior data efficiency relative to iMiner. (b) We compare two different user prompts: Sars2Pro and Sars2Pro+Ro5. All results were generated with 2000 valid SMILES at a temperature of $T=$1.0 and a maximum of 128 new tokens.}
\label{fig:3d}
\end{figure}
\noindent
Figure \ref{fig:3d}a compares the docking scores of the original iMiner algorithm against SmileyLlama as a function of epoch number and with number of generated molecules per iteration. We notice first an improved data efficiency compared to iMiner, in which SmileyLlama requires only $\sim$25\% of the epochs to reach a similar level of improved docking score. Furthermore iMiner's diversity crashes with more epochs, which explains the sharpening peaks in later iterations (Figure \ref{fig:3d}) and quantified further in Supplementary Figure 2 against the Guacamol benchmarks. This simply reflects convergence in the docking score, i.e. there are fewer novel molecules as docking score reaches the highest values. By contrast, SmileyLlama maintains more diversity while greatly improving the docking score to iMiner (Figure \ref{fig:3d}a) with minor degradation in validity compared to iMiner (Supplementary Figure 5). 

Figure \ref{fig:3d}b shows the property distributions of the final optimized set of novel molecules from SmileyLlama from the above prompt. While the property distributions are satisfactory for the number of hydrogen bond donors and acceptors, the MW and logP results are not conforming to drug-like values. This indicates some inadequacy of the iMiner reward function, such that the CLM would require a reweighting and/or new terms in the loss/reward function, other hyperparameter tuning, and/or expensive retraining. But a unique advantage of SmileyLlama is that the distribution of generated molecules’ properties can be shifted using nothing more than prompt engineering, with no retraining required. Figure \ref{fig:3d}b shows that combining prompts such as \textbf{Output a SMILES string for a drug like molecule with the following properties: High SARS2PRO, <= 5 H-bond donors, <= 10 H-bond acceptors, <= 500 molecular weight, <= 5 logP} (\textbf{High SARS2Pro+Ro5}) improves properties such as MW and logP and drug-likeness scores substantially with some expected loss in high docking scores since smaller molecules make fewer intermolecular interactions. 
\begin{figure}[H]
\centering
\includegraphics[width=0.99\textwidth]{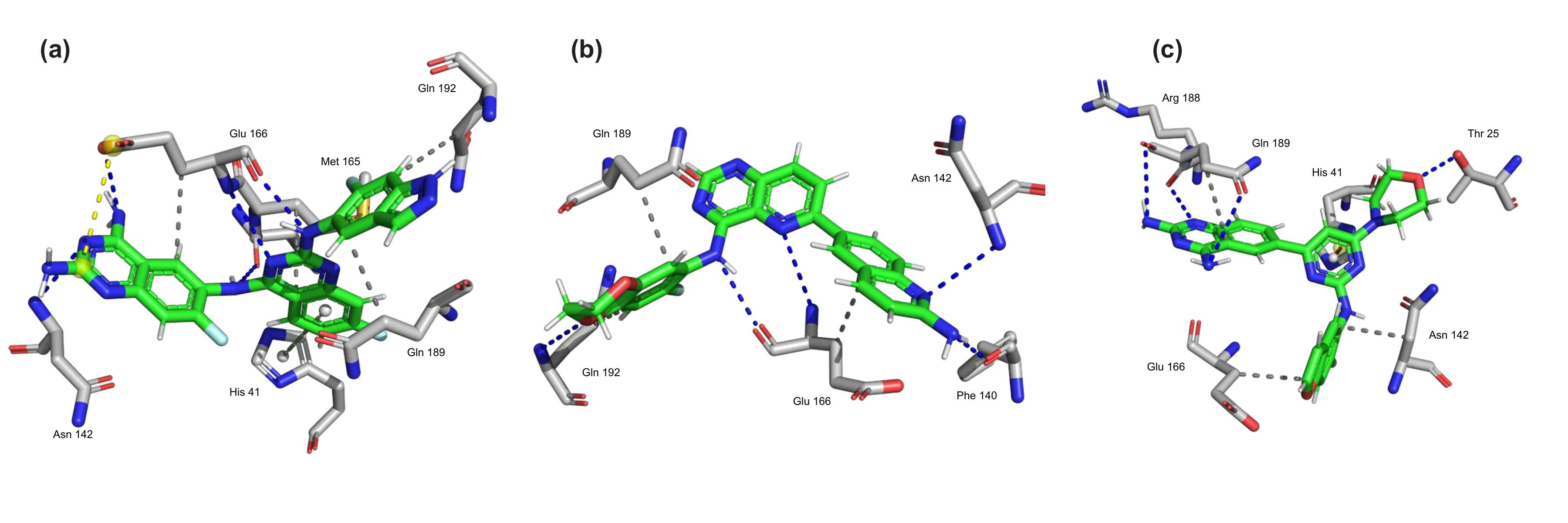}
\vspace{-3mm}
\caption{\textit{SmileyLlama de novo generated molecules in the active site of SARS2 main protease.} Surface rendering of the SmileyLlama generated molecules in the SARS2 Mpro canonical binding pocket. Generated by SmileyLlama after optimization with (a) the \textbf{SARS2PRO} prompt. (b) and (c) the \textbf{SARS2Pro+Ro5} prompt. Supplementary Data file molecule\_examples.csv provides their SMILES string and docking scores, and Supplementary Figure 6 shows the docking pose for some of the highest scoring ligands. Blue is nitrogen, red is oxygen, green indicates ligand carbons and light grey indicates residue carbons.}
\label{fig:molecules}
\end{figure}

Figure \ref{fig:molecules} and Supplementary Figure 6 provides a set of novel molecules from SmileyLlama docked in the MPro active site with the two engineered prompts \textbf{High SARS2Pro} and \textbf{High SARS2Pro+Ro5}. Two of the higher scoring molecules resemble the variations of the perampanel drug with the trefoil structure which are tested inhibitors optimized by the Jorgensen group.\cite{zhang2021} However, unlike the molecules from their study that consistently retained the central pyridinone ring\cite{zhang2021}, SmileyLlama molecules have replaced the trefoil hub with the pyrimidine functional group (see Figure \ref{fig:molecules}c). Higher docking scores are found for quite different drug scaffolds (Figure \ref{fig:molecules}(a,b)), but in all cases there is no significant homology match found in the Therapeutic Target Database.\cite{Zhou2024} This would indicate that the generative capabilities of SmileyLlama are robust and outside of the pre-trained Llama model. Finally, the proposed drugs are synthetically accessible\cite{Ertl2009}, as made clear from their on average SA$\sim$ 3. Precise details can be found in Supplementary data file example\_molecules.csv.

\subsection{SmileyLlama Outside of Chemical Language Modeling}
While SFT and DPO alters Llama in the creation of SmileyLlama, we find that SmileyLlama can still converse in English if it is prompted to do so, and some sample conversations are included in the Methods Section 4.6. As a more quantitative measure of its residual abilities, we evaluate performance of SmileyLlama on the Language Model Evaluation Harness containing the MMLU GPQA, Math-Hard, and MMLU-Pro benchmarks\cite{eval-harness, mmlu, wang2024mmlupro, rein2023gpqa, hendrycks2021math}. Supplementary Table 3 and Supplementary Figure 7 show that SmileyLlama generally performs worse on moral scenarios, and amusingly SmileyLlama also performs worse on subjects dealing with chemistry than Llama. This in part is likely due to the tendency for SmileyLlama to complete prompts relating to chemistry with a SMILES string. In addition, accuracy errors in the MMLU tests have also been noted recently\cite{gema2025mmlu}, and thus SmileyLlama's degraded performance in chemistry may also be in part an artifact of poor testing benchmarks. Overall this result is somewhat encouraging, since it implies the possibility that LLM-derived CLMs can inherit and take advantage of the natural language processing ability of their foundation model. SmileyLlama already does this--- we can steer the properties of the molecules it generates and the chemical space it explores using natural language prompts while still retaining some ability to process non-chemical natural language. However, more work will be required to create SmileyLlama as an additional capability of an LLM, which might be achievable with larger parameter foundational LLMs.

\section{Discussion}
Our study clarifies a few crucial points for CLMs derived from LLMs going forward. First, it is not necessary to pretrain a specialized model on chemistry-specific text to generate molecules from a text description; a much less resource-intensive SFT training run on prompt-following on a dataset of a few million molecules with a commodity LLM is enough to do that. Second, DPO provides another resource-efficient way of optimizing the model to produce molecules that score well on a targeted objective without needing in-context examples, instead relying on the generative nature of the model itself for good and bad examples. A corollary to this is the finding that SmileyLlama can combine its knowledge gained during single-objective optimization to perform well at a task specifying multiple objectives, elicited by combining the prompts (as opposed to requiring training on both prompts), which is a welcome outcome. Even so, there are still limitations to and tradeoffs within the SmileyLlama framework and within our investigation for drug discovery.  Additional factors for good drug candidates must also inhibit "off-target effects" and/or be robust to mutation of the protein or virus among other downstream requirements. While SmileyLlama was not explicitly optimized for generating molecules with these qualities in this work, the DPO framework laid out here should be extensible to optimizing molecules for these characteristics. Even so, while DPO improves adherence to the prompt, it does so at the cost of narrowing the distribution of properties or diversity, which may not be desirable in all application areas or early stages of discovery. Furthermore, SmileyLlama still struggles in data-poor regimes, for instance in the task of generating macrocycles.

The prompting and optimization framework for modifying LLMs to explore chemical space broadly or to narrow the search to specific regions shown here could also be leveraged for molecular design outside of drug discovery, such as the use of SMILES for elaborating on transition metal complexes\cite{Rasmussen2025}. One could also imagine that casting a chemical problem as a linguistic construct could also allow for other applications, such as our recent work on chemical synthesis\cite{Sun2025}. As with many of the fields touched by LLMs this decade, the newly opened frontier of possibility in chemistry is as vast as it is exciting.

\section{Methods}

\subsection{Details of properties for fine-tuning}

\textbf{Overview of selected properties for fine-tuning}. When fine-tuning Llama to generate drug-like molecules, we carefully assess various design choices and proceed with the following properties, emphasizing those that medicinal chemists would consider when proposing \textit{de novo} drug molecules. We categorized and summarized all 12 properties into 4 subgroups as follows.
\begin{itemize}
    \item \textit{Physiochemical properties}. Absorption, distribution, metabolism, and excretion (ADME) are the crucial criteria to quantify the localization and concentration of drug molecules within the body after administration. As a result, we build on the list of properties proposed in the classical Lipinski's rule of 5 \cite{LIPINSKI20013} with some modern additions like TPSA to generate drug-like molecules that could demonstrate decent ADME.
    \begin{itemize}
        \item Number of Hydrogen Bond Donors (\#HBD)
        \item Number of Hydrogen Bond Acceptors (\#HBA)
        \item Molecular Weight (MW)
        \item Log of Partition Coefficient (LogP)
        \item Topological Polar Surface Area (TPSA)
        \item Fraction of $sp^3$ hybridized carbon atoms ($Fsp^3$)
    \end{itemize}
    \item \textit{Structure flexibility features}. Binding sites within a targeted biomolecule (most often a protein) display by nature complex 3D geometry, with key potential sites of drug-target interactions (amino acid side chains, as an example) somewhat fixed in space. The protein, however, has a dynamic structure and even the binding pocket undergoes changes in shape. Drug-like molecules need to be sufficiently rigid to make efficient interactions with the target protein, including in most cases a high degree of selectivity over making corresponding interactions with related proteins).  Perhaps less intuitive is that drug-like molecules must be flexible enough to maintain those interactions as the protein adapts its conformation. There is a "Goldilocks principle" at play, where too rigid or too flexible are each undesired extremes. Here, we chose the following two properties to account for the flexibility aspect.
    \begin{itemize}
        \item Number of Rotatable Bonds (\#rot)
        \item Whether the molecule contains a macrocycle (defined as an 8-membered ring or larger)
    \end{itemize}
    \item \textit{Pattern-based features}. In practical drug discovery, there are always some key patterns and/or scaffolds that medicinal chemists would like to hold onto or get rid of. For instance, in the lead optimization phase, retaining the key moiety and desired chemical formula are rather essential. Meanwhile, avoiding chemically unstable groups, PAINS molecules \cite{baell_chemistry_2014}, and molecules that would cause structure alerts could increase the chance of success in development. Therefore, we have the following three properties for fine-tuning.
    \begin{itemize}
        \item Avoidance of undesirable chemical patterns
        \item Retention of specified substructure (between 50 and 250 Da in molecular weight)
        \item Chemical formula
    \end{itemize}
    \item \textit{Covalent warhead feature}. Drugs can be broadly categorized into noncovalent and covalent drugs, depending on whether the drug reacts with its target. That is, an electrophilic group of a covalent inhibitor might form a bond with a nucleophilic amino acid side chain of its target protein. The reactive functional group of a covalent inhibitor is called a warhead. While most drugs are non-covalent, either can be desired. To give the model the ability to generate covalent binders, we also curated common covalent warhead-related SMARTS patterns from the Enamine fragment library \cite{enamine_essential_libraries} to indicate whether our generated molecules have the capacity to covalently bind to the target or not.
    \begin{itemize}
        \item Whether the molecule contains common covalent warhead-related SMARTS patterns, and which of these patterns appear in the molecule
    \end{itemize}
\end{itemize}

\noindent
\textbf{Prompting options used in fine-tuning}. To incorporate the properties mentioned above into the training, we used several ways of prompting to satisfy the requirement from target uses. 

For numerical properties, including all physiochemical properties and \#rot, we prompted Llama by providing a specific range that the training molecules falls into for that specific category. All the cutoff values used for ranges are either commonly used standards in drug discovery or derived from the training distribution. Besides the range guidance, we added the prompt that tell Llama exactly how many \#HBDs and \#HBAs are contained in the training data that gives its ability to do more nuanced generation. If a property falls into multiple valid ranges---for instance, 4 H-bond donors is less than or equal to all of 4, 5 and 7---we select one of the ranges at random to place in the prompt (if the properly is picked to be placed into the prompt). It is important that the list of ranges for a property span all possible molecules, otherwise a prompt which omits information biases the model toward producing molecules with values of that property outside of the list of ranges. If we never included information in the prompt about when a molecule had \textit{over} 7 H-bond donors, but did sometimes include information about the number of H-bond donors when they number under or equal to 7, then omitting the information would give the model a \textit{hint} that the number of H-bond donors is more likely to be over 7. Doing this during training would bias results during inference. This is the same reason we specify undesirable properties such as the presence of bad SMARTS strings in the prompt sometimes. If the random number generator decided that a prompt should contain a substructure but the SMILES in question did not have any BRICS substructures, we added "no BRICS substructure" to the list of properties in lieu of a substructure.

For other categorical properties, we used a combination of RDKit modules, SMILES strings, and SMILES arbitrary target specification (SMARTS) strings to recognize if certain properties or chemical patterns are present in the training input. Unlike the objective of containing the scaffold exactly, chemical pattern avoidance and covalent warhead recognition required matching of more general substructures and/or certain functional groups. Here, we used SMARTS strings as our representation because of its ability of matching chemical patterns. More details about the specific SMARTS patterns used are shown later in this section.

Below is a detailed list of possible components of that could appear in a training prompt.
\begin{itemize}
    \item $N$ H-bond donors, $N = \leq3, \leq4, \leq5, \leq7, >7$
    \item $N$ H-bond acceptors, $N = \leq3, \leq4, \leq5, \leq10, \leq15, >15$
    \item $N$ Molecular weight, $N = \leq300, \leq400, \leq500, \leq600, >600$
    \item $N$ LogP, $N = \leq3, \leq4, \leq5, \leq6, >6$
    \item $N$ Rotatable bonds, $N = \leq 7, \leq10, >10$
    \item $N$ Fraction $sp^3$, $N = <0.4, >0.4, >0.5, >0.6$
    \item $N$ TPSA, $N = \leq90, \leq140, \leq200, >200$
    \item a macrocycle, no macrocycles
    \item has bad SMARTS, lacks bad SMARTS
    \item has covalent warheads, lacks covalent warheads
    \item substructure of *a\textunderscore smiles\textunderscore string*
    \item a chemical formula of *formula*
\end{itemize}

\noindent
\textbf{SMART patterns used to identify bad chemical groups}. Li \textit{et al.} pointed out a list of bad chemical patterns that exists in ChEMBL database, which will negatively affect compound generation\cite{iminer}. In this work, we used the same list of SMARTS patterns as their work to avoid bad patterns, including cyclopentadiene, cyclopentadiene ylidenes, aromaticity-breaking tautomers, antiaromatic system, unstable halogen-heteroatom bonds, unstable fused rings, allenic system, thiazyl linkages, and peroxide bonds. In Supplementary Table 2, we also present the frequency of sampling undesirable chemical groups in ChEMBL and across different generative models.
\begin{itemize}
    \item\verb|[C^2]1=[C^2]-[C^2]=[C^2]~[C;!d4]~[C;!^2;d2]1|
    \item\verb|[C^2]1~[C^2]~[C^2]~[C^2]~[C;!^2;d2]~[N]1|
    \item\verb|[#6^2]1~[#6^2]~[#6^3;!d4]~[#6^2]2~[#6^2]~| \\\verb|[#6^2]~[#6^2]~[#6^2](~[*])~[#6^2]~2~[#6^2]~1|
    \item\verb|[#6]1(=[*])[#6]=[#6][#6]=[#6]1|
    \item\verb|[#6]1=[#6][R{2-}]=[R{2-}]1|
    \item\verb|[#6^2]1~[#6^2]~[#6^2]~[#6^2]~[#6^1]~[#6^1]~1|
    \item\verb|[#7,#8,#16]-[#9,#17,#35,#53]|
    \item\verb|[r3,r4]@[r5,r6]|
    \item\verb|[*]=[#6,#7,#8]=[*]|
    \item\verb|[#7,#16]=[#16]|
    \item\verb|[#8]-[#8]|
\end{itemize}
In addition to the patterns mentioned above, we use the following SMARTS patterns to enforce our generated pyrroles to be one of the following correct forms.
\begin{itemize}
    \item\verb|[N^2]1~[C,N;^2](=[*])~[C,N;^2]~[C,N;^2]~[C^3]1|
    \item\verb|[N^2]1~[C,N;^2]~[C,N;^2](=[*])~[C,N;^2]~[C;^3]1|
    \item\verb|[N^2]1~[C,N;^2]~[C,N;^2]~[C,N;^2](=[*])~[C;^3]1|
    \item\verb|[C,N;^2](=[*])1~[N;^2]~[C,N;^2]~[C,N;^2]~[C;^3]1|
    \item\verb|[C,N;^2]1~[N;^2]~[C,N;^2](=[*])~[C,N;^2]~[C;^3]1|
    \item\verb|[C,N;^2]1~[N;^2]~[C,N;^2]~[C,N;^2](=[*])~[C;^3]1|
\end{itemize}

\noindent
\textbf{SMART patterns used to encode common covalent warhead-related functional groups}. Common covalent warheads are extracted from the Enamine Covalent Screening and Covalent Fragment Library \cite{enamine_essential_libraries}. The list of SMARTS strings is shown below.

\begin{itemize}
\item sulfonyl fluorides: \verb|[#16](=[#8])(=[#8])-[#9]|
\item chloroacetamides: \verb|[#8]=[#6](-[#6]-[#17])-[#7]|
\item cyanoacrylamides: \verb|[#7]-[#6](=[#8])-[#6](-[#6]#[#7])=[#6]|
\item epoxides: \verb|[#6]1-[#6]-[#8]-1|
\item aziridines: \verb|[#6]1-[#6]-[#7]-1|
\item disulfides: \verb|[#16]-[#16]|
\item aldehydes: \verb|[#6](=[#8])-[#1]|
\item vinyl sulfones: \verb|[#6]=[#6]-[#16](=[#8])(=[#8])-[#7]|
\item boronic acids/esters: \verb|[#6]-[#5](-[#8])-[#8]|
\item acrylamides: \verb|[#6]=[#6]-[#6](=[#8])-[#7]|
\item cyanamides: \verb|[#6]-[#7](-[#6]#[#7])-[#6]|
\item chloroFluoroAcetamides: \verb|[#7]-[#6](=[#8])-[#6](-[#9])-[#17]|
\item butynamides: \verb|[#6]#[#6]-[#6](=[#8])-[#7]-[#6]|
\item chloropropionamides: \verb|[#7]-[#6](=[#8])-[#6](-[#6])-[#17]|
\item fluorosulfates: \verb|[#8]=[#16](=[#8])(-[#9])-[#8]|
\item beta lactams: \verb|[#7]1-[#6]-[#6]-[#6]-1=[#8]|
\end{itemize}

In section 3.2, we investigate SmileyLlama's performance on the following 387 tasks, grouped into the families for which the averages are shown in the table.

\begin{itemize}
\item exactly k H-bond donors, from k=0 to k=5
\item exactly k H-bond acceptors, from k=0 to k=10
\item <= k H-bond donors, for k=3,4,5,7
\item <= k H-bond acceptors, for k=3,4,5,10,15
\item <= k Molecular weight, for k=300,400,500,600
\item <= k LogP, for k=3,4,5,6
\item <= 7, <=10, >10 Rotatable bonds
\item > 0.4, > 0.5, > 0.6, < 0.4 Fraction sp3
\item <=90, <=140, <=200 TPSA
\item a macrocycle
\item no macrocycles
\item has bad SMARTS (not shown in table but included for completeness)
\item lacks bad SMARTS
\item lacks covalent warheads
\item has covalent warheads (one for each of the 16 covalent warheads in the section above)
\item a substructure of (one of each of 320 Enamine fragments \cite{enamine_essential_libraries})
\item \verb|<= 5 H-bond donors, <= 10 H-bond acceptors, <= 500 Molecular weight, <= 5 LogP|
\item \verb|<= 3 H-bond donors, <= 3 H-bond acceptors, <= 300 Molecular weight, <= 3 LogP|
\end{itemize}

\subsection{Prompt formats and examples.}
We assess the ability of Llama to generate SMILES strings as a baseline. Below are examples of system and user prompts to illustrate the methods we used to prompt Llama and SmileyLlama. The Llama prompts are built using the Llama instruct prompting format, while the SmileyLlama, robotic prompt and blank prompts are using the alpaca format to reproduce what was used in the most recent supervised fine-tune of the foundation model.

For the case of Llama zero-shot, we use the following format, with no pre-filled responses, when generating data for the guacamol benchmark. We chose to use a user prompt asking for "no other output" since in our informal experiments, Llama would often respond indirectly, including english text discussing SMILES strings without this explicit instruction to only generate SMILES strings.
\newline\texttt{
System prompt:\\ \textbf{You love and excel at generating SMILES strings of drug-like molecules}\\
User prompt:\\  \textbf{Please generate a drug-like smiles string and no other output:}}
\newline\newline
Llama-k-shot has k pre-filled responses using ChEMBL molecules. In this example, we will show the system prompt and user prompt with three pre-filled ChEMBL molecules.:
\newline\texttt{
System prompt:\\ \textbf{You love and excel at generating SMILES strings of drug-like molecules}\\
User prompt:\\  \textbf{Please generate a drug-like smiles string and no other output:}\\
Response:\\  \textbf{Cc1cc(c(C)n1CCOC)C(=O)CSc1nc2nc(cc(n2n1)C)C}\\
User prompt:\\  \textbf{Please generate a drug-like smiles string and no other output:}\\
Response:\\  \textbf{N(c1nc([C@]23N=C(N)SC[C@@H]3C[C@H](C)OC2)sc1)C(c1ncc(nc1)OC)=O}\\
User prompt:\\  \textbf{Please generate a drug-like smiles string and no other output:}\\
Response:\\  \textbf{c1nn([C@H](C(NCCc2sccc2C)=O)CC)cc1}\\
User prompt:\\  \textbf{Please generate a drug-like smiles string and no other output:}}
\newline\newline
\noindent
Moving on to prompts used for supervised fine-tuning and subsequent inference, we first give system and user prompts for SmileyLlama. Because the user prompt is dependent on whether any properties are selected to be specified, we give both versions here. The user prompt with no properties should sample from a distribution most similat to ChEMBL, so we use this format when sampling SMILES for assessment of the GuacaMol Benchmark.
\newline\texttt{
System prompt:}\\ \textbf{"You love and excel at generating SMILES strings of drug-like molecules"}\\
User prompt (no properties selected):\\ \textbf{\"Output a SMILES string for a drug-like molecule:"}\\
User prompt (properties selected):\\ \textbf{Output a SMILES string for a drug-like molecule with the following properties: <property 1>, <property 2>, <property 3>, ..."}
\newline\newline
\noindent
Below are the system and user prompt used in the "robotic prompt" control of prompt phrasing for Guacamol inference. It should be noted that the SFT dataset for this "robotic prompt" control had the same user prompts as SmileyLlama (including specified properties), but a system prompt of \texttt{Generate a SMILES string of a drug-like molecule according to the user's input:}
\newline\texttt{
System prompt:\\ \textbf{"}Generate a SMILES string of a drug-like molecule according to the user's input:"}\\
User prompt (no properties selected):\\ \textbf{"Output a SMILES string for a drug-like molecule:"}\\
User prompt (properties selected):\\ \textbf{Output a SMILES string for a drug-like molecule with the following properties: <property 1>, <property 2>, <property 3>, ..."}
\newline\newline
\noindent
Below are the system an user prompts for the "Blank prompt" example in Table 1.
\newline\texttt{
System prompt:}\\ \textbf{""}\\
User prompt (no properties selected):\\ \textbf{""}\\
User prompt (properties selected):\\ \textbf{<property 1>, <property 2>, <property 3>, ..."}

\subsection{Additional Training Details} 
We performed both SFT and DPO on Llama using the Axolotl Package~\cite{noauthor_axolotl-ai-cloudaxolotl_nodate}. For both SFT and DPO, we use the Low-Rank Adaptation (LoRA) applied to the linear layers of the model and FlashAttention with an Adam optimizer, cross-entropy loss, and a cosine learning rate scheduler with a maximum learning rate of $2 \times 10^{-4}$ for SFT and $2 \times 10^{-5}$ for DPO.\cite{hu_lora_2021,dao2023flashattention2} All prompts were formatted according to the Alpaca instruction format \cite{alpaca}. Additional parameters for our training are a LoRA rank of 32, a LoRA alpha of 16, a LoRA dropout of 5\%, and 10 warmup steps. We inherit these hyperparameters from standard practice with Axolotl, such as LoRA hyperparameters identical to these were used in the Hermes 3 SFT\cite{Hermes3TechnicalReport}. For SFT, we trained for 1 epoch using a batch size of 64 on a single 4xA40 node for approximately 32 hours with a validation set $\sim$5\% the size of the original, an amount which would cost approximately \$53 in October 2025 on Vast.ai. We also note that we randomized the SMILES string representation of each molecule, and we tokenized all SMILES strings with the Llama3 tokenizer\cite{dubey_llama_2024} when interfacing with SmileyLlama.

We used the HuggingFace Transformers\cite{hftransformers} library to perform inference. Unless otherwise indicated, we used a temperature of 1.0. To avoid biasing generations, we do not restrict the possible tokens produced at any particular step by setting top\_p or top\_k. We allow a maximum of 128 new tokens, which truncates the size of the generated SMILES strings, noting that larger values of this hyperparameter lead to generally similar results, with the exception of a broader, less drug-like distribution of molecules after several iterations of optimization for MPro binding within the iMiner framework. We also note that SmileyLlama can fall victim to the repeat curse; on occasion SmileyLlama will continue producing tokens indefinitely with some repetitive pattern on sufficiently long molecules, emphasizing the need for a token cutoff.

\subsection{GuacaMol Benchmark definitions}
The GuacaMol benchmark assesses generative models based on five metrics.\cite{brown2019guacamol}
\begin{itemize}
    \item Validity: The proportion of the first N generated strings which are RDKit-parsable and have more than 0 atoms.
    \item Uniqueness: The number of distinct molecules in a set of N total valid generated strings, divided by N.
    \item Novelty: The number of N valid, unique generated strings which do not represent a molecule in the training set
    \item KL-Divergence: The distribution of a variety of physiochemical descriptors are calculated for both the generated molecules and the training set, and their similarity is assessed through KL-divergence.
    \item Frechet ChemNet Distance: The Frechet Distance between the distribution of generated molecules' activations on a neural network called ChemNet's penultimate layer and the training set's activations.
\end{itemize}

All benchmarks were performed with 10,000 samples at a temperature T =1.0 and a maximum of 128 new tokens for Llama and SmileyLlama. We note that, due to the proprietary nature of Llama’s training data and the SMILES contained therein, our assessments of Llama zero-shot’s novelty with respect to ChEMBL is not as meaningful as the novelty assessment of SmileyLlama and the CLMs. Finally, we note that FCDGuac=exp(-0.2*FCD) compresses the FCD distances themselves for the convenience of defining a 0 to 1 value like the other Guacamol metrics\cite{brown2019guacamol}. Hence the scores should actually be considered from a log perspective such that an FCD below 5 (i.e. FCDGuac =0.37) is in strong agreement with distributions of drug-like properties. Only when the FCDGuac score drops by one to two orders of magnitude are they considered to be poor FCD scores. We recommend in the future that others report the more straightforward FCD distances themselves to avoid this interpretative confusion. 

\subsection{iMiner reinforcement learning with SmileyLlama}
The iMiner generative model uses an Average Stochastic Gradient Descent Weigh-Dropped Long Short-Term Memory (AWD-LSTM)\cite{iminer} recurrent neural network that predicts probability of string tokens to concatenate on a molecular string representation until a complete molecule is generated. In the subsequent RL stage, 2000 molecules in each epoch (typically $\sim$ 50 epochs) are sent to AutoDock simulations in parallel, and the docking scores are circled back to the RL model to adjust its parameters so that molecules generated in the next iteration will have better scores while retaining drug-likeness. Given the ascendancy of attention mechanisms and transformer architectures, such as those inherent in SmileyLlama, we replaced the generative AWD-LSTM component of iMiner with SmileyLlama, and replaced the iMiner optimization algorithm, Proximal Policy Optimization (PPO)\cite{iminer} with DPO. This is largely due to the high memory requirement of PPO; since DPO does not need to fit a reward model to the predicted and realized rewards it requires far less memory than PPO. This becomes even more true when tuning a large model compared to a small one, since generally the reward model trained is the same size as the language model used to generate strings.

We use a scoring function of three times the docking score plus the iMiner drug-likeness function\cite{iminer} to score all molecules per iteration. The drug-likeliness score is an extension of the widely used quantitative estimate of drug-likelihood (QED) value\cite{iminer}, and is defined as:

\begin{align}
     S_{DL}(X)=\sum_{i}{\sigma_i \log{p_i(\mathrm{prop}_i(X))}}
\end{align}

\noindent
where $\mathrm{prop}_i(X)$ calculates the $i$th property of a molecule $X$ and $p_i$ is defined by the probability distribution of property $i$ by all molecules in the ChEMBL database. The parameter $\sigma_i$ is defined as:
\begin{align}
    \sigma_i=S_i^{-1}/\sum_j{S_j^{-1}}
\end{align}
where $S_i$ is the entropy of the distribution of property $i$,
\begin{align}
    S_i=-\sum_x{p_i(x) \log{p_i(x)}}
\end{align}
such that a narrower distribution from the ChEMBL database contribute more to the drug likeliness score, and defines the weights for each property as proportional to the inverse of the entropy. Invalid molecules were assigned a score of -10. More details can be found in the original iMiner work\cite{iminer}.

After every 2000 distinct molecules are generated per iteration using SmileyLlama, we score each molecule and randomly pair it with another for the DPO procedure. The molecule with the higher score is designated as the "winner" and the other as the "loser", and we construct a dataset with eight such pairings for each molecule and then optimize SmileyLlama using DPO with respect to this dataset. We empirically found 8 pairs to work well; more than 8 pairs tended to lead to a lack of diversity, whereas fewer than 8 pairs resulted in less sample efficient optimization. This process is described in Supplementary Information Algorithm 3. 

\subsection{SmileyLlama Outside of Chemical Language Modeling}
While SFT clearly alters the model, one might wonder whether SmileyLlama retains LLM abilities unrelated to exploring chemical space. We find that SmileyLlama inherits some ability to process natural language, however, SmileyLlama will sometimes respond to chemistry-related queries with a SMILES string rather than a sensible, natural-language answer. Below are four sample responses from SmileyLlama to four prompts. Each was generated with greedy decoding, which is equivalent to sampling at a temperature of 0. All responses were generated using the Alpaca instruction format, with a system prompt/instruction of ``You are a helpful assistant''. As shown below, SmileyLlama is capable of writing working python code for simple functions and answering some questions in natural language. This demonstrates that, while SmileyLlama was designed to be a CLM, it still inherits some of Llama's natural language processing ability. However, this is not perfect, as SmileyLlama will sometimes respond to chemistry-related queries with a SMILES string, rather than the correct answer.\par

\noindent\texttt{Input: What is the molecular weight of hydrazine?\\
Output: \textbf{N1N=C(C)C(C1)C}}
\newline
(Note: This is the SMILES representation of 3,4-dimethyl 2-pyrazoline, rather than hydrazine)
\newline
\newline
\noindent\texttt{Input: What is a Diels-Alder reaction in organic chemistry?\\
Output: \textbf{A Diels-Alder reaction is a [4+2] cycloaddition between a diene and a dienophile.}}
\newline

\noindent\texttt{Input: Who was Christopher A. Lipinski in the field of chemistry? What was his "Rule"?.\\
Output: \textbf{Christopher A. Lipinski was a chemist who developed the "Lipinski's Rule of Five" which states that for a drug to be orally bioavailable, it should have the following properties: \\
1) C(C)Cn1c2c(cccc2)c(c1)C(Nc1ccccc1)=O \\
2) A molecular weight of 500 or less \\
3) No more than 5 H-bond donors \\
4) No more than 10 H-bond acceptors}}

An example of code generation is given in the Supplementary Figure 8 
Furthermore, we used the Language Model Evaluation Harness to test SmileyLlama's performance on the MMLU GPQA, Math-Hard, wikitext, and MMLU-Pro benchmarks \cite{eval-harness, wang2024mmlupro, rein2023gpqa, hendrycks2021math}. The results in Supplementary Table 3 and Supplementary Figure 7 show that SmileyLlama generally performs only a little worse on most tasks compared to Llama. For practical purposes, the SFT procedure for creating SmileyLlama seems to have created additional functionality to Llama, and we expect that much larger LLMs will be even better at sustaining their original capabilities.

\textbf{Unphysical requests}. Since SmileyLlama can take in prompts purely in natural language, it is possible to give SmileyLlama a nonsensical instruction. Here, we ask SmileyLlama to generate molecules which are impossible, in principle, to generate. In both cases we investigate, SmileyLlama doesn't refuse the request. Rather, it will generate SMILES strings not satisfying the (impossible) conditions.

First, we ask SmileyLlama to generate a molecule which has two properties in contradiction with each other: a substructure of 1,2,3,4-tetramethoxybenzene and 3 or fewer H-bond acceptors. This is impossible since 1,2,3,4-tetramethoxybenzene has four H-bond acceptors itself, and so any molecule with this as a substructure must have at least that many H-bond acceptors. With greedy decoding, SmileyLlama gives the following molecule:
\newline\newline
\noindent\texttt{Input: Output a SMILES string for a drug like molecule with the following\\ properties: a substructure of c1c(OC)c(OC)c(OC)c(OC)c1, <= 3 H-bond acceptors:\\
Output: \textbf{c1c(c(c(c(c1)OC)OC)OC)C1C2C(CCC=1)C1C(CCC=2)C1}}
\newline\newline
This molecule has a substructure of 1,2,3-trimethoxybenzene, which is similar but not identical to 1,2,3,4-tetramethoxybenzene. However, the output molecule has 3 H-bond acceptors, which obeys the second instruction in the prompt.
\newline\newline\newline
We can also ask for a blatantly unphysical molecule, such as one with negative six H-bond donors:
\newline\newline
\texttt{Input: Output a SMILES string for a drug like molecule with the\\\indent\indent following properties: <= -6 H-bond donors:\\
Output: \textbf{c1c2c(ccc1)C(=O)N(C2=O)CC(Nc1ccc(cc1)C(=O)Nc1ccc(cc1)C(Nc1ccc(cc1)\\\indent\indent C(Nc1ccc(cc1)C(Nc1ccc(cc1)C(Nc1ccc(cc1)C(Nc1ccc(cc1)C(Nc1ccc(cc1)\\\indent\indent C(Nc1ccc(cc1)C(Nc1ccc(cc1)C(Nc1ccc(cc1)C(Nc1ccc(cc1)C(Nc1ccc(cc1)C(N
}}
\newline\newline
This cut off at 128 new tokens, which was the maximum for this particular experiment. To our knowledge, it will repeat this pattern indefinitely, which we have tested for up to 8192 new tokens, a pattern not new to LLMs.

This demonstration of SmileyLlama's behavior is not exhaustive, there are likely to be many other requests which elicit responses different from what a naïve extrapolation of these anecdotes would indicate. Here, we simply aimed to give examples of some of the interesting behavior that SmileyLlama can have when asked to perform tasks outside of the drug discovery purposes for which it was originally grown.

\section*{Data Availability}
\noindent
The data included here contains all SMILES strings we used from ChEMBL (chembl.txt), the dataset containing the BRICS substructures of these ChEMBL molecules, the dataset containing prompts and responses used for SFT to create SmileyLlama from Llama, and the dataset used to create SmileyLlama-DPO-opt from SmileyLlama. Source data for all Figures is available with this manuscript. Data used for this study can be found in references\cite{SmileyLlamaData,SmileyLlamaData2}

\section*{Code Availability}
\noindent
Code used for this study can be found in reference \cite{SmileyLlamaCode} and models can be found in references \cite{SmileyLlamaModel1, SmileyLlamaModel2, SmileyLlamaModel3}. 

\section*{Acknowledgments}
This work was supported in part by the National Institute of Allergy and Infectious Disease grant U19-AI171954 for the drug molecule application. We thank the CPIMS program, Office of Science, Office of Basic Energy Sciences, Chemical Sciences Division of the U.S. Department of Energy under Contract DE-AC02-05CH11231 for support of the machine learning. We thank Riza Özçelik for kindly providing the retraining code for different CLMs for benchmarking and Nicole Kennedy for suggesting properties of molecules useful to medicinal chemists.

\section*{Author Contributions Statement}
\noindent
J.M.C.,K.S., A.G., and T.H.G. defined goals and designed the project. J.M.C., K.S., A.G., and D.B. carried out the optimizations. J.M.C., K.S., A.G., and T.H.G. wrote the paper. All authors discussed the results and made comments and edits to the manuscript.

\section*{Competing Interests Statement}
\noindent
The authors declare no competing interests.

\clearpage
\bibliography{library}
\bibliographystyle{naturemag}

%\pagebreak
%%%%%%%%%% Merge with supplemental materials %%%%%%%%%%
%%%%%%%%%% Prefix a "S" to all equations, figures, tables and reset the counter %%%%%%%%%%

%\include{si}
\end{document}

\section*{Table of Contents}
\textbf{Algorithms used for SFT and DPO...............................................................................................................2}

\noindent
\textbf{Supplementary Tables 1-3..................................................................................................................3}

\noindent
\textbf{Supplementary Figures 1 and 2................................................................................................................4}

\noindent
\textbf{Supplementary Figure 3..............................................................................................................................5}

\noindent
\textbf{Supplementary Figure 4..............................................................................................................................6}

\noindent
\textbf{Supplementary Figure 5..............................................................................................................................7}

\noindent
\textbf{Supplementary Figure 6..............................................................................................................................8}

\noindent
\textbf{Supplementary Figure 7..............................................................................................................................9}

\noindent
\textbf{Supplementary Figure 8.............................................................................................................................10}

\newpage

\section{Algorithms used for SFT and DPO}

\begin{algorithm}[H]
\small
\caption{Pseudocode for generating the SFT dataset}
\begin{algorithmic}
\FORALL{SMILES\_STRING $\in$ ChEMBL}
    \STATE PROMPT $\gets$ "Output a SMILES string for a drug like molecule with the following properties:"
    \FORALL{(PROPERTY\_FUNCTION, PROPERTY\_NAME) $\in$ shuffle(PROPERTIES)}
        \IF{random\_coin\_flip() $=$ HEADS}
            \STATE VALUE $\gets$ PROPERTY\_FUNCTION(SMILES\_STRING)
            \STATE VALUE\_RANGE $=$ RANDOM\_RANGE\_VALUE\_FALLS\_IN(VALUE)
            \STATE PROMPT $\gets$ PROMPT $+$ PROPERTY\_NAME $+$ " of " $+$ VALUE\_RANGE $+$ ", "
        \ENDIF
    \ENDFOR
    \STATE prompts.append(PROMPT)
    \STATE completions.append(SMILES\_STRING)
\ENDFOR
\end{algorithmic}
\end{algorithm}

\begin{algorithm}[H]
\small
\caption{Pseudocode for generating the DPO dataset for improving prompt obedience}
\begin{algorithmic}
\FORALL{(SMILES\_STRING, CONDITIONS) $\in$ RESPONSES}
    \IF{MEETS\_CONDITIONS(SMILES\_STRING, CONDITIONS) $=$ True}
        \STATE winners.append(SMILES\_STRING)
    \ELSE
        \STATE losers.append(SMILES\_STRING)
    \ENDIF
\ENDFOR
\STATE pairs $\gets$ randomly\_pair(winners, losers)
\STATE \textbf{assert:} $|\text{pairs}| = \min\{|\text{winners}|, |\text{losers}|\}$
\end{algorithmic}
\end{algorithm}

\begin{algorithm}[H]
\small
\caption{Pseudocode for generating the DPO dataset for Mpro optimization every iteration}
\begin{algorithmic}
\STATE PROMPT $\gets$ "Output a SMILES string for a drug like molecule with the following properties: High SARS2PRO"
\WHILE{$|\text{SMILES\_LIST}| < \text{num\_responses}$}
    \STATE SMILES\_LIST $\gets$ SmileyLlama.inference(PROMPT, num\_responses)
    \STATE SMILES\_LIST $\gets$ SMILES\_LIST.remove\_redundancies()
\ENDWHILE

\FOR{SMILES $\in$ SMILES\_LIST}
    \STATE iMiner.compute\_score(SMILES)
\ENDFOR

\FOR{i $=$ 1 to 8}
    \FOR{SMILES $\in$ SMILES\_LIST}
        \WHILE{True}
            \STATE RANDOM\_OTHER\_SMILES $\gets$ pick\_random(SMILES\_LIST)
            \IF{iMiner.retrieve\_score(SMILES) $>$ iMiner.retrieve\_score(RANDOM\_OTHER\_SMILES)}
                \STATE DATASET.append((PROMPT, SMILES, RANDOM\_OTHER\_SMILES))
                \STATE \textbf{break}
            \ELSIF{iMiner.retrieve\_score(SMILES) $<$ iMiner.retrieve\_score(RANDOM\_OTHER\_SMILES)}
                \STATE DATASET.append((PROMPT, RANDOM\_OTHER\_SMILES, SMILES))
                \STATE \textbf{break}
            \ENDIF
        \ENDWHILE
    \ENDFOR
\ENDFOR
\end{algorithmic}
\end{algorithm}

\clearpage

\clearpage

\section{\fontsize{16}{16}\selectfont Supporting Tables}

\begin{table}[h]
    \centering
    \caption{\textit{GuacaMol benchmarks comparing SmileyLlama to related models, ablations, and a model trained with a different prompt format.} Similar to Table 1 in the main text, we generate molecules at a temperature $T=$1.0 and a maximum of 128 new tokens. These benchmark scores suggest that the SmileyLlama SFT procedure outlined here can be extended towards adapting other LLMs into CLMs, including Qwen and smaller variants within the Llama 3 Herd. We also note that changing the prompt templates, as shown in the "robotic phrasing" and "blank prompt" rows, does not appear to significantly affect these results.}
    \small
    \vspace{-2mm}
    \begin{tabular}{lcccccc}
        \hline\hline
        \textbf{Benchmark} & \textbf{Validity} & \textbf{Uniqueness} & \textbf{Novelty} & \textbf{KLdiv$_{Guac}$} & \textbf{FCD$_{Guac}$} \\ \hline
        \textbf{SmileyLlama} & 0.958 & 1.000 & 0.987 & 0.967 & 0.686 \\        %\textbf{SmileyLlama} & 0.960 & 1.000 & 0.988 & 0.970 & 0.553 \\%Got rid of canonical smiles
        \textbf{robotic phrasing} & 0.952 & 1.000 & 0.988 & 0.963 & 0.696 \\
       \textbf{blank prompt} & 0.953 & 1.000 & 0.984 & 0.966 & 0.676 \\
        \textbf{SmileyLlama-1B} & 0.911 & 1.000 & 0.987 & 0.968 & 0.588 \\
        \textbf{SmileyLlama-3B} & 0.929 & 1.000 & 0.989 & 0.973 & 0.617 \\
        \textbf{SmileyQwen2.5-7B} & 0.964 & 1.000 & 0.987 & 0.908 & 0.516 \\
        \hline
    \end{tabular}
\label{tab:SIguacamol}
\end{table}

\begin{table}[ht]
\centering
\caption{\textit{Frequency of generating or sampling undesirable chemical structures in ChEMBL, SmileyLlama, and other SOTA methods.} Here, we randomly sample 10,000 molecules from the ChEMBL dataset, SmileyLlama and our retrained models to compare their statistics of five different categories of undesirable chemical patterns.}
    \begin{tabular}{lccccc}
    \hline
    \textbf{Model} & ChEMBL & SmileyLlama & LSTM & GPT & S4 \\
    \hline
    Unsaturated Benzene     & 0.55\%    & 0.1 \%	    & 0.04\%	& 0.04\%	&  0.03\% \\
    Unsaturated Naphthalene & 0.02\%    & 0.01\%   	& 0.02\%	& 0.01\% & 0.01\% \\
    Wrong Pyrrole           & 0.29\%    & 0.02\%    & 0.02\%	& 0.03\%	& 0.04\% \\
    Cyclopentadiene Ylidene & 0.96\%    & 0.1 \%    & 0.04\%	&0.06\%	 & 0.05\% \\
    Benzyne                 & 0.01\%    & 0.0 \% 	& 0.0\%	    &0.0\%	&0.0\%\\
    \hline \hline
    \end{tabular}
\label{tab:SMART}
\end{table}

\begin{table}[h!]
    \centering
    \caption{\textit{Performance of Meta-Llama-3.1-8B-Instruct versus SmileyLlama in a variety of LLM benchmarks.} We report the mean score, assessed with the exact\_match scoring method for all tasks besides perplexity.}
    \small
        \begin{tabular}{lccccc}
    \hline
    \textbf{Model} & Llama-3.1-8B-Instruct & SmileyLlama \\
    \hline
    Wikitext (word perplexity) ($\downarrow$) & 8.6401 & 8.8934\\
    MMLU (0-shot) ($\uparrow$) & 0.68 & 0.65 \\
    GPQA (chain of thought) ($\uparrow$) & 0.30 & 0.038\\
    Math-Hard ($\uparrow$) & 0.24&	0.039 \\
    MMLU-Pro ($\uparrow$) & 0.47& 0.063 \\
    \hline \hline
    \end{tabular}
\label{tab:properties_specification}
\end{table}

\clearpage
\section{\fontsize{16}{16}\selectfont Supporting Figures}

\begin{figure} [H]
\centering
\includegraphics[width=0.95\textwidth]{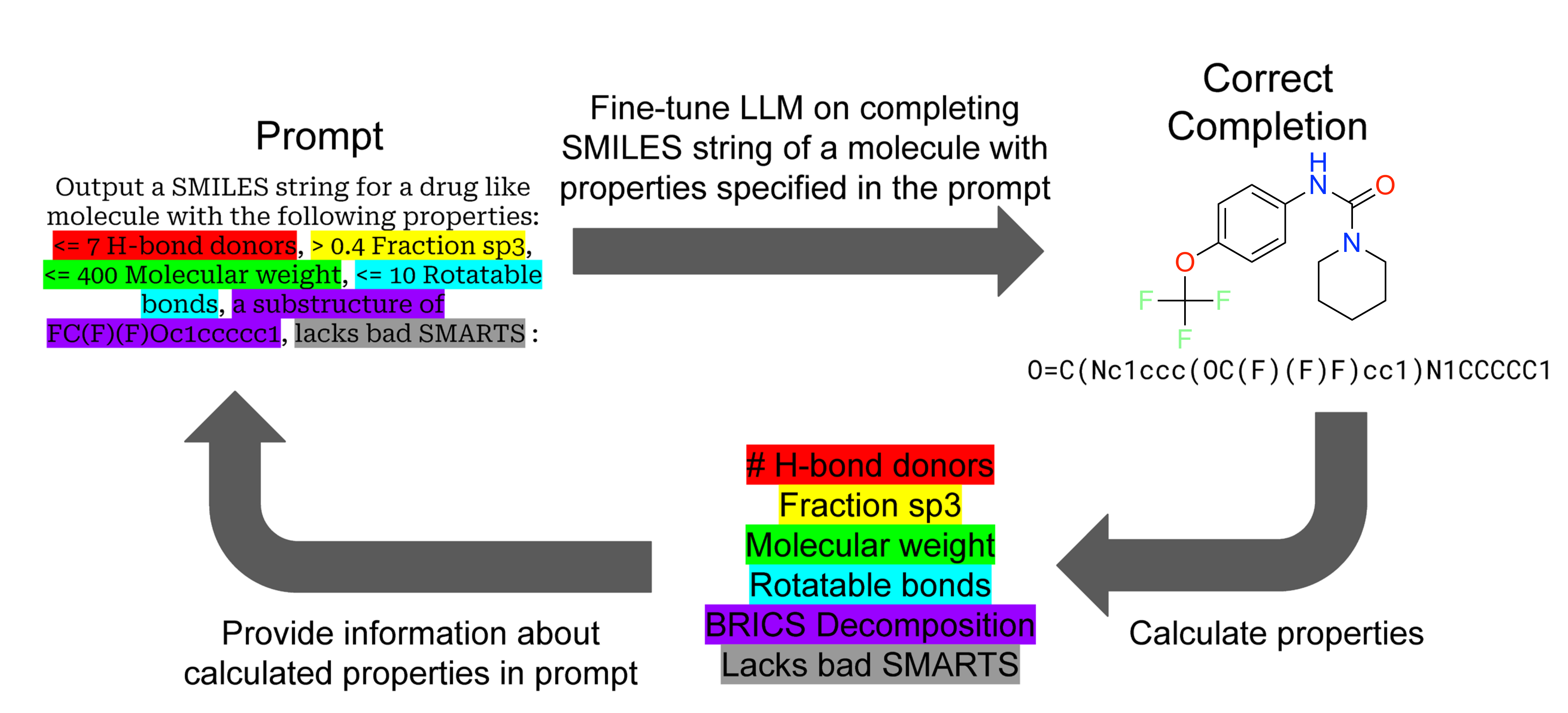}
\vspace{-2mm}
\caption{\textit{A visualization of the SFT workflow for Smiley-Llama.} Given the Llama-3.1-8B-Instruct model~\citep{dubey_llama_2024}, we used prompt-response pairs consisting of calculated molecular properties and completed SMILES strings to fine-tune Llama on SMILES strings completions, yielding SmileyLlama. Crucially, we construct the prompt for each example using properties calculated from the correct response (a SMILES string from ChEMBLv33).}
\label{fig:sft_workflow}
\end{figure}

\begin{figure}[h!]
\centering
\includegraphics[width=0.75\textwidth]{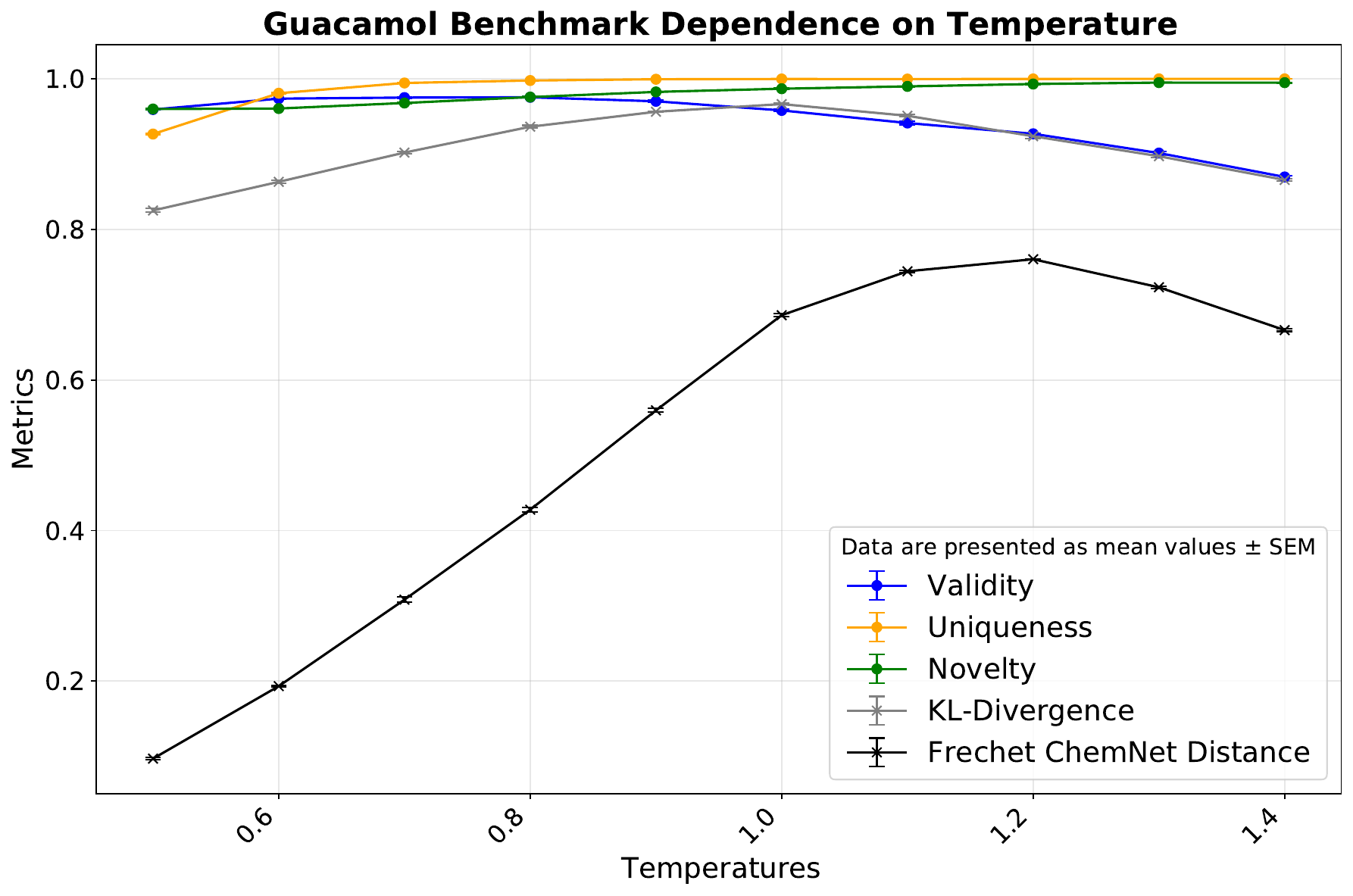}
\caption{\textit{Guacamol benchmark results as a function of temperature.} We used n=3 runs to assess how validity, uniqueness and novelty depends on temperature, a parameter which affects the randomness of outputs generated during inference. Error bars are defined by the standard error. Temperature can be increased during inference for more random molecules, sacrificing validity for diversity and novelty. We generally use a temperature of T=1.0 during this study.}
\label{fig:mmlu}
\end{figure}

\begin{figure}[h!]
\centering
\includegraphics[width=0.95\textwidth]{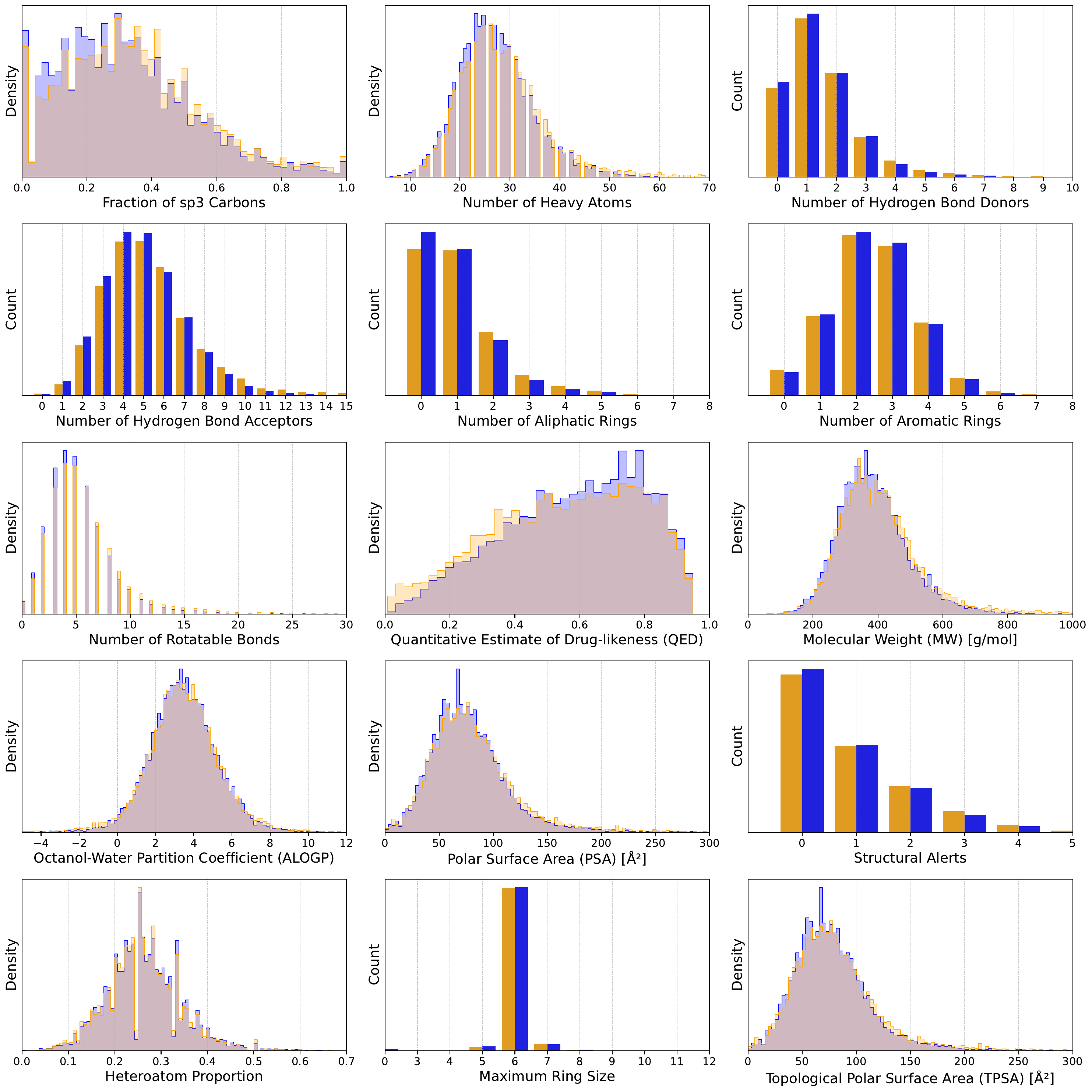}
\caption{\textit{Distribution comparisons for different properties of the generated molecules from the LSTM CLM (blue) with molecules from the training dataset from ChEMBL (gold).} To be compared with Figure 1b in the main text and Figure S4.}
\label{fig:lstm_properties}
\end{figure}

\begin{figure}[h!]
\centering
\includegraphics[width=0.95\textwidth]{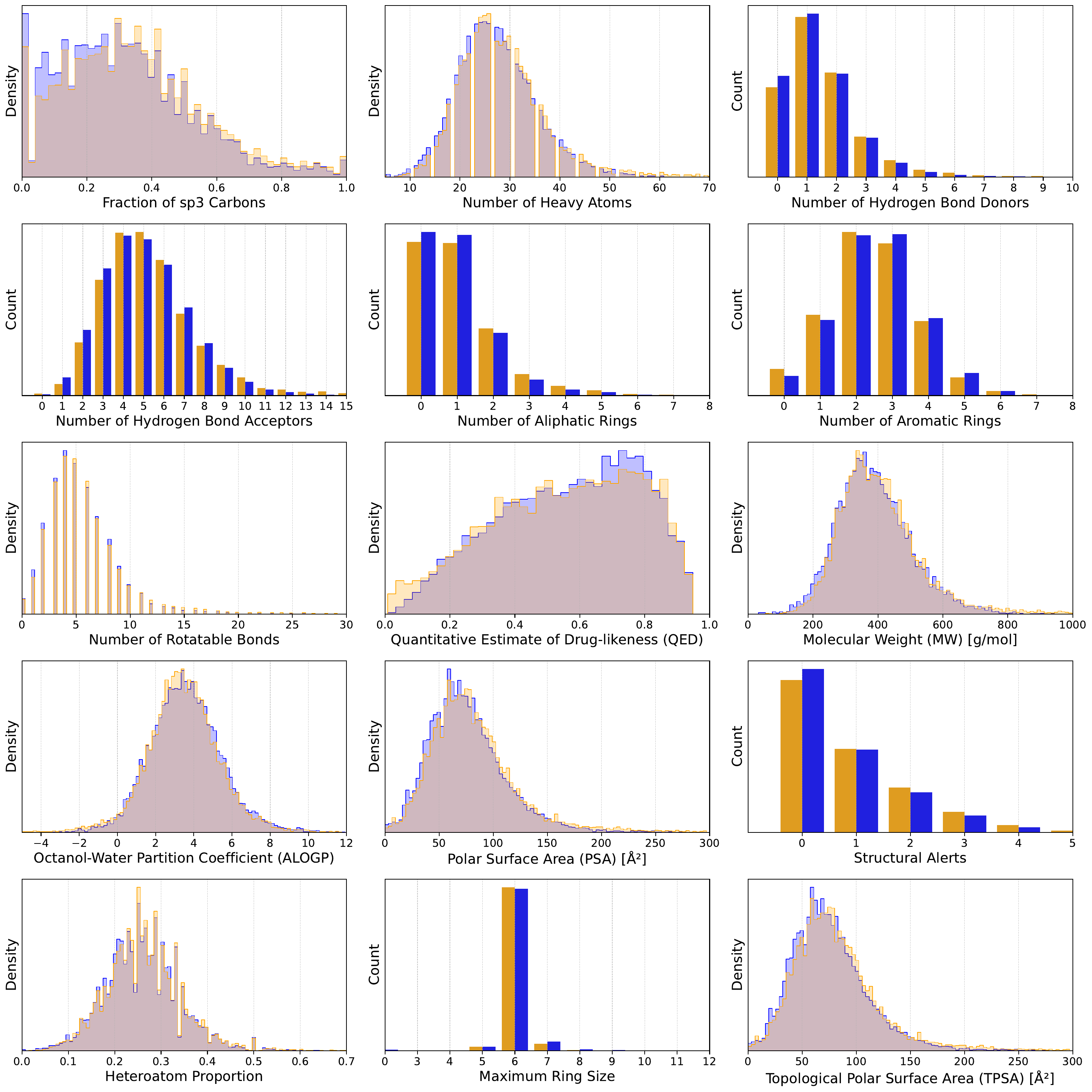}
\caption{\textit{Distribution comparisons for different properties of the generated molecules from the GPT model (blue) with molecules from the training dataset from ChEMBL (gold).} To be compared with Figure 1b in the main text and Figure S3.}
\label{fig:gpt_properties}
\end{figure}

\begin{figure}[h!]
\centering
\includegraphics[width=0.95\textwidth]{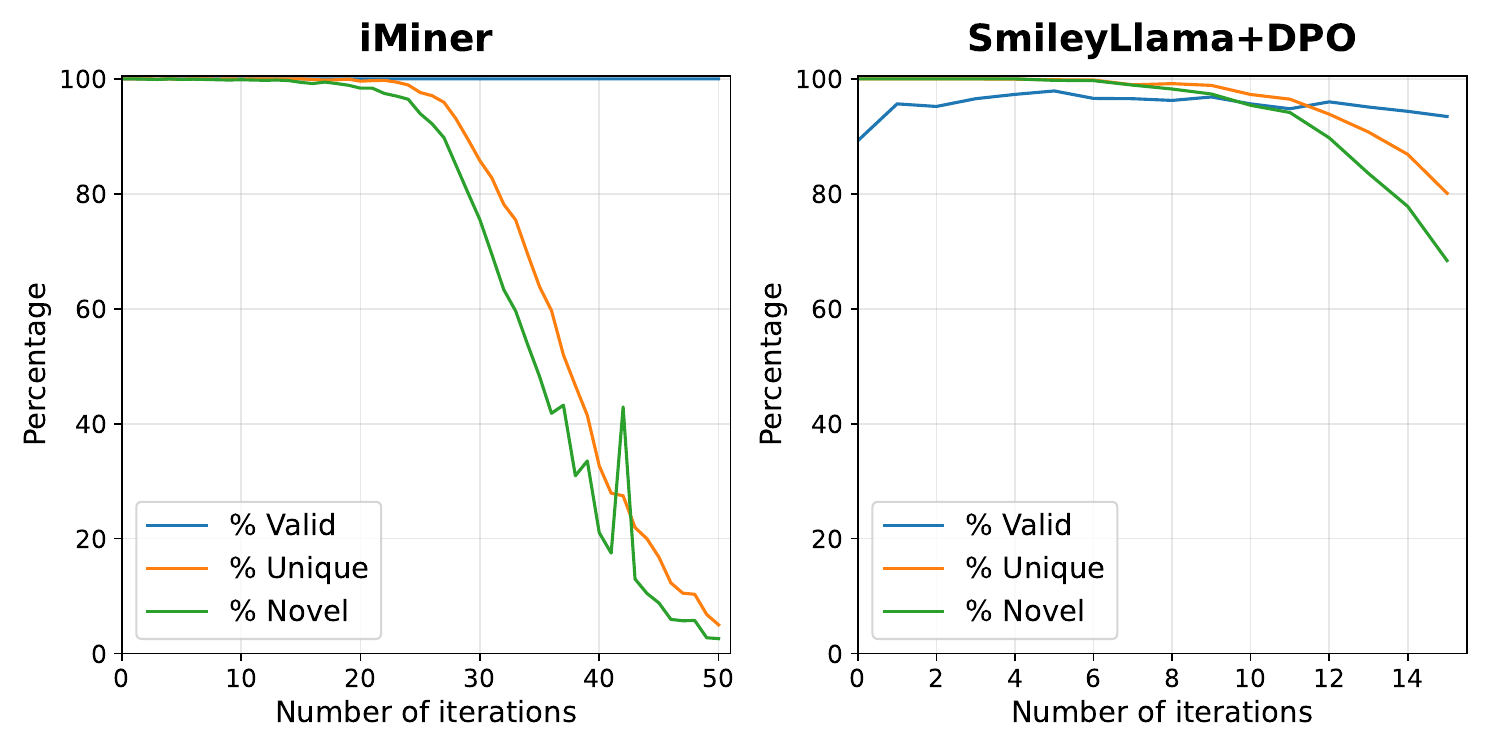}
\caption{\textit{Validity, uniqueness and novelty over optimization for SARS2-Mpro inhibition} We plot the validity (fraction of molecules which are valid), uniqueness (number of distinct molecules) and novelty (fraction of molecules which have not been generated before) of molecules sampled from the generative models of iMiner with 2000 molecules per epoch, and for SmileyLlama with DPO (SL+DPO) with 2000 and 1000 molecules per epoch. Note that novelty is not defined with respect to ChEMBL; here it's a measure of how good the model is at generating molecules that it has not generated in a previous iteration.}
\label{fig:mmlu}
\end{figure}

\begin{figure}[h!]
\centering
\includegraphics[width=0.99\textwidth]{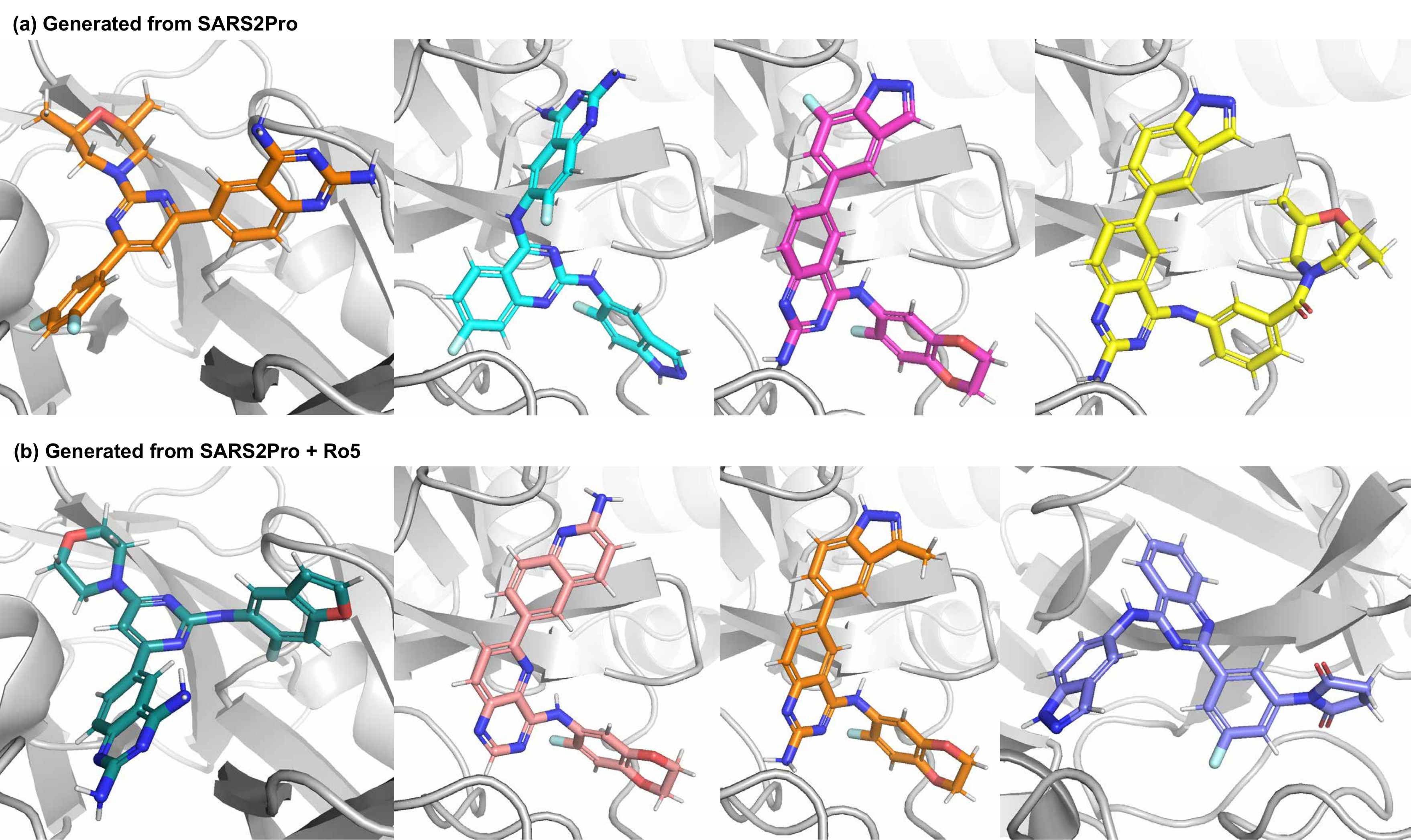}
\vspace{-3mm}
\caption{\textit{SmileyLlama de novo generated molecules in the active site of SARS2 main protease.} Surface rendering of the SmileyLlama generated molecules in the SARS2 Mpro canonical binding pocket. Generated by SmileyLlama after optimization with the \textbf{SARS2PRO} and \textbf{SARS2Pro+Ro5} prompts. Supplementary Table S4 provides their SMILES string and docking scores. Blue is nitrogen, white is hydrogen, red is oxygen, otherwise carbon}
\label{fig:molecules}
\end{figure}

\begin{figure}[h!]
\centering
\includegraphics[width=0.98\textwidth]{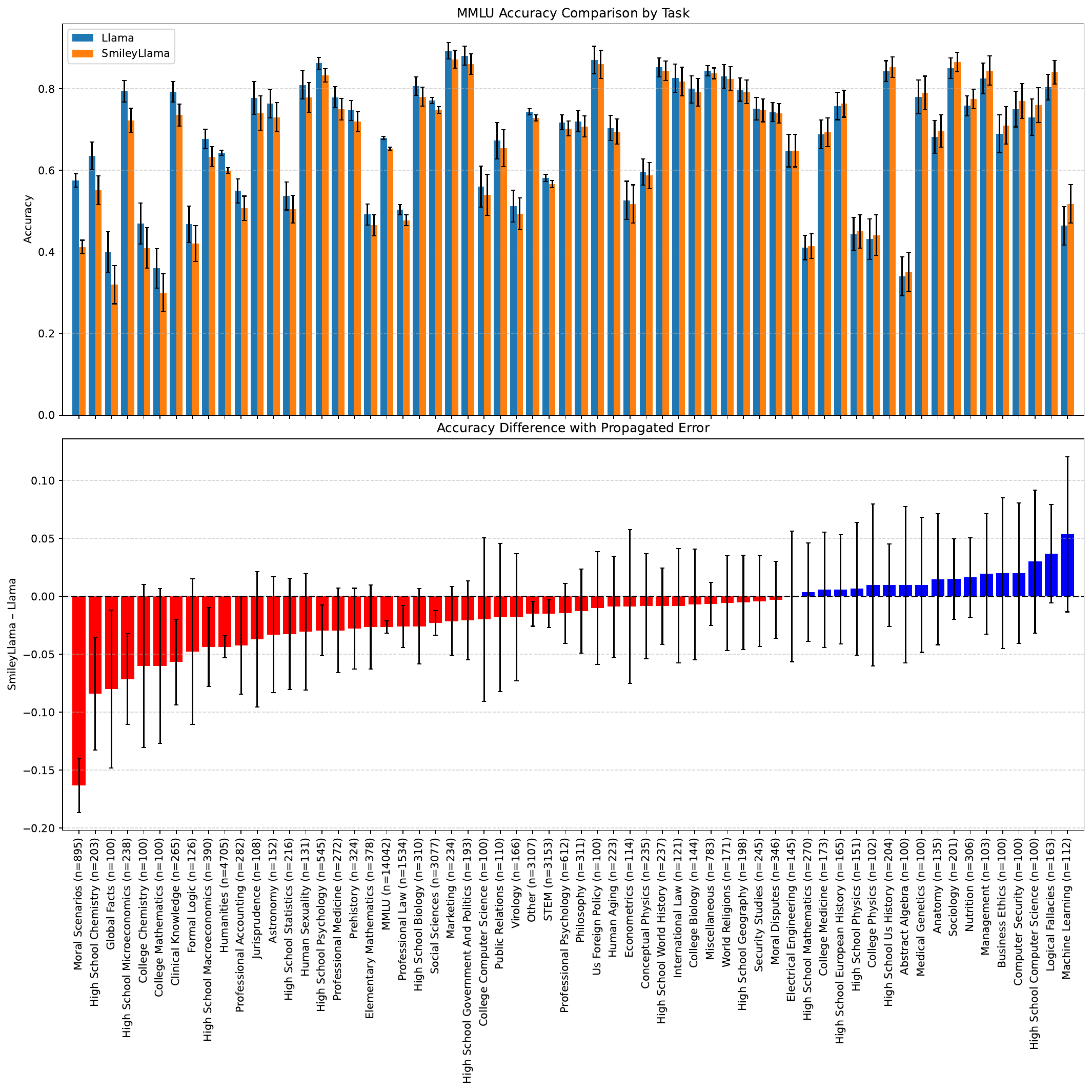}
\caption{\textit{A visualization of how the SmileyLlama SFT procedure affects zero-shot Massive Multitask Language Understanding (MMLU) score\cite{mmlu}. There were no technical replicates performed, since the accuracy was assessed through loglikelihoods for each answer, which are deterministic. All bars are mean, and error bars are standard error (top) and propagated standard error (bottom) } (top) Generally, MMLU scores decrease with SFT. One might expect that fine-tuning which improves the ability to generate molecules may translate to improved performance on chemistry, but this was found to be the opposite of the case; chemistry-specific sections of the MMLU saw among the worst degradation in score after SFT. Anecdotally, SmileyLlama will often respond to a chemistry-related query with a SMILES string. The "moral scenarios" section of the MMLU had the largest difference in score between the base and instruct Llama models; while SmileyLlama scores about 41\% compared to Llama-3.1-8B Instruct's 57\%, the base model Llama-3.1-8B score centers around 28\%. This may be due to much of the "moral knowledge" coming during post-training. On the other hand, there are some tasks, such as "Machine Learning", where performance appears to improve, although the standard error bars for each of these tasks include the mean so we cannot conclusively say that the SFT procedure used for SmileyLlama improved Llama's abilities outside of SMILES string generation. Error bars are given as standard error in the upper figure and propagated standard error in the second, respectively.}
\label{fig:mmlu}
\end{figure}

\begin{figure}[h!]
\caption{\textit{Example of code generation}.}
    \noindent\texttt{Input: Write me some python code which calculates the n'th Fibonacci number}\\
\noindent\texttt{Output:}
\begin{lstlisting}
def fibonacci(n):
    if n <= 0:
        return "Input should be a positive integer"
    elif n == 1:
        return 0
    elif n == 2:
        return 1
    else:
        a, b = 0, 1
        for _ in range(2, n):
            a, b = b, a + b
        return b
\end{lstlisting}
\label{fig:mmlu}
\end{figure}

\clearpage
\bibliography{library}
\bibliographystyle{naturemag}

%This defines the bibliographies style. Search online for a list of available styles.
%\bibliographystyle{abbrv}